\documentclass[12pt,preprint]{aastex}
\usepackage[version=3]{mhchem}
\usepackage{lscape}
\usepackage{longtable}

\makeatletter
\newcounter{reaction}
\renewcommand\thereaction{C\,\arabic{reaction}}
\newcommand\reactiontag{\refstepcounter{reaction}\tag{\thereaction}}
\newcommand\reaction@[2][]{\begin{equation}\ce{#2}%
\ifx\@empty#1\@empty\else\label{#1}\fi%
\reactiontag\end{equation}}
\newcommand\reaction@nonumber[1]{\begin{equation*}\ce{#1}%
\end{equation*}}
\newcommand\reaction{\@ifstar{\reaction@nonumber}{\reaction@}}
\makeatother

\shorttitle{Photochemistry of Terrestrial Exoplanet Atmospheres}
\shortauthors{Hu et al.}

\begin{document}

\title{Photochemistry in Terrestrial Exoplanet Atmospheres II: \ce{H2S} and \ce{SO2} Photochemistry in Anoxic Atmospheres}

\author{Renyu Hu$^1$, Sara Seager$^{1,2}$, William Bains$^{1,3}$}
\affil{$^1$Department of Earth, Atmospheric and Planetary Sciences, Massachusetts Institute of Technology, Cambridge, MA 02139}
\affil{$^2$Department of Physics, Massachusetts Institute of Technology, Cambridge, MA 02139}
\affil{$^3$Rufus Scientific, Melbourn, Royston, Herts, United Kingdom}
\email{hury@mit.edu}

\begin{abstract}
Sulfur gases are common components in the volcanic and biological emission on Earth, and are expected to be important input gases for atmospheres on terrestrial exoplanets. We study the atmospheric composition and the spectra of terrestrial exoplanets with sulfur compounds (i.e., \ce{H2S} and \ce{SO2}) emitted from their surfaces. We use a comprehensive one-dimensional photochemistry model and radiative transfer model to investigate the sulfur chemistry in atmospheres ranging from reducing to oxidizing. The most important finding is that both \ce{H2S} and \ce{SO2} are chemically short-lived in virtually all types of atmospheres on terrestrial exoplanets, based on models of \ce{H2}, \ce{N2}, and \ce{CO2} atmospheres. This implies that direct detection of surface sulfur emission is unlikely, as their surface emission rates need to be extremely high ($>1000$ times Earth's volcanic sulfur emission) for these gases to build up to a detectable level. We also find that sulfur compounds emitted from the surface lead to photochemical formation of elemental sulfur and sulfuric acid in the atmosphere, which would condense to form aerosols if saturated. For terrestrial exoplanets in the habitable zone of Sun-like stars or M stars, Earth-like sulfur emission rates result in optically thick haze composed of elemental sulfur in reducing \ce{H2}-dominated atmospheres for a wide range of particle diameters (0.1 - 1 $\mu$m), which is assumed as a free parameter in our simulations. In oxidized atmospheres composed of \ce{N2} and \ce{CO2}, optically thick haze, composed of elemental sulfur aerosols (\ce{S8}) or sulfuric acid aerosols (\ce{H2SO4}), will form if the surface sulfur emission is 2 orders of magnitude more than the volcanic sulfur emission of Earth. Although direct detection of \ce{H2S} and \ce{SO2} by their spectral features is unlikely, their emission might be inferred by observing aerosol-related features in reflected light with future generation space telescopes.
\end{abstract}


\section{Introduction}

A large number of super Earths have been detected by radial velocity surveys and transit surveys. Super Earths are exoplanets with masses no more than 10 times the mass of Earth. The atmospheres of super Earths are important because characterization of super Earth atmospheres is a substantial step towards eventually characterizing truly Earth-like exoplanets. Attempts to observe super Earth atmospheres are growing (e.g., Batahla et al. 2011 for Kepler 10 b; Demory et al. 2012 and Ehrenreich et al. 2012 for 55 Cnc e), and one super Earth/mini Neptune GJ 1214b is being observed in as much detail as possible (e.g., Bean et al. 2010; Croll et al. 2011; D\'esert et al. 2011; Berta et al. 2012; De Mooij et al. 2012). In particular, the transmission spectra of GJ 1214b is nearly flat from 0.6 to 5 $\mu$m, which has ruled out a planet with an \ce{H2}-dominated cloud-free atmosphere. The observational push to super Earth characterization has the potential to provide a handful of super Earth atmospheres to study in the coming years. In the more distant future, the community still holds hope that a direct-imaging space-based mission under the Terrestrial Planet Finder concept will come to existence, with planets and their atmospheres observed in reflected light.

The concentration of trace gases in super Earth atmospheres is controlled by the component gas emission from the surface and subsequent sinks in the atmosphere (chemical reactions initiated by UV photolysis). Even with a trace amount, some gases may leave significant footprints in the planet's spectra, for instance \ce{H2O}, \ce{CO2}, and \ce{O3} in Earth's atmosphere. In contrast to giant exoplanets where the atmospheric composition is mainly controlled by the elemental abundance, the steady-state composition of super Earth atmospheres is mainly controlled by photochemical processes. We have developed a comprehensive photochemistry model that computes chemical compositions of any terrestrial exoplanet atmospheres, ranging from reducing to oxidizing, with all key non-equilibrium processes taken into consideration, including photolysis, chemical kinetics, vertical diffusion of molecules, atmospheric escape, dry and wet deposition, and condensation and sedimentation of concern condensable species (Hu et al.  2012; referred to as Paper I hereafter). The effects of surface gas emission on different super Earth atmospheres  can be investigated with the photochemistry model.

Sulfur gases emitted from the surface and their photochemical products significantly shape the spectra of rocky bodies in the Solar System. The most striking feature of Venus' atmosphere is a high planetary albedo due to thick \ce{H2SO4} clouds. Photochemistry models of the Venusian atmosphere have been developed and the formation of \ce{H2SO4} in the dry \ce{CO2}-dominated atmosphere have been simulated (e.g., Yung \& DeMore 1982; Zhang et al. 2012; Krasnopolsky 2012). The latest photochemistry model of Venus' atmosphere has assumed a constant mixing ratio\footnote{Mixing ratio is defined as the ratio of the amount of a gas in a given volume to the total amount of all gaseous constituents in that volume.} of \ce{SO2} (i.e., $\sim10$ ppm) at the altitude of 47 km (implying significant source of \ce{SO2} from below) and predicted the formation of \ce{H2SO4} at the altitudes around 66 km (Zhang et al. 2012; Krasnopolsky 2012). Io, the innermost moon of Jupiter, is believed to have very intensive and variable volcanic activity with \ce{SO2} emission, and the Io's atmosphere is dominated by photolysis of \ce{SO2} and subsequent formation and polymerization of elemental sulfur (e.g., Moses et al. 2002). On early Earth, the sulfur chemistry may have been very different from now, featuring the formation of sulfur aerosols as well as sulfate aerosols as the atmosphere was anoxic (Kasting et al. 1989; Pavlov \& Kasting 2002; Kasting \& Catling 2003; Zahnle et al. 2006). The formation of insoluble sulfur aerosols is believed to be critical for the record of mass independent fractionation that timed the rise of oxygen in Earth's atmosphere (e.g., Farquhar et al, 2000; Pavlov \& Kasting 2002; Zahnle et al. 2006). In addition, organosulfur compounds, such as dimethyl sulfide (DMS) and methanethiol (\ce{CH3SH}), have also been suggested to be biosignatures of the early Earth (Pilcher 2003). The greenhouse effect of \ce{SO2} has been suggested to have contributed to the warming of early Mars (e.g., Halevy et al. 2007), a proposition that has been challenged by photochemistry studies that predict sulfur and sulfate aerosol formation as a result of \ce{SO2} emission on early Mars and the anti-greenhouse effect of these aerosols (Tian et al. 2010).

Terrestrial exoplanets could have a wide range of sulfur gas emission. Sulfur gases are common volcanic gases on Earth, and in some scenarios may be more prevalent on exoplanets. Sulfur is a major building block for rocky planets and the abundance of sulfur is about one sixteenth that of carbon in the Solar System (Lodders 2003). On present-day Earth, sulfur compounds, mainly in the form of \ce{H2S} and \ce{SO2}, are dominant volcanic gases in additional to hydrogen, oxygen and carbon emissions. In the Earth's magma, the dissolved sulfur has a weight percentage ranging from 10$^{-4}$ to 10$^{-2}$ (e.g., Wallace \& Edmonds 2011) and degassing of sulfur compounds as the magma decompresses when rising to the surface provides a global volcanic sulfur flux of about $3\times10^{9}$ S cm$^{-2}$ s$^{-1}$ (Seinfeld \& Pandis 2006). 90\% of current Earth's sulfur emission is in the form of \ce{SO2}, whereas the \ce{H2S}/\ce{SO2} ratios for individual volcanoes vary widely between 0.01 and 1 (see Holland 2002 and references therein). An intriguing fact is that the amounts of sulfur compounds in Earth's atmosphere is extremely low despite the substantial emission rates, due to very short chemical timescales. On Earth, the lifetime of \ce{H2S} and \ce{SO2} in the troposphere is only 2 days, which makes the mixing ratio of these gases in the atmosphere very small (Seinfeld \& Pandis 2006). In Earth's troposphere, the main sink of \ce{H2S} is the reaction with the hydroxyl radical \ce{OH} (Lelieveld et al. 1997), and the main sink of \ce{SO2} is the removal by dry and wet deposition.

Another reason that \ce{H2S} and \ce{SO2} photochemistry is interesting is that Earth-based biological processes involve sulfur compounds. There are multiple ways that life can produce \ce{H2S}, including the reduction of sulfate (e.g., Watts 2000) and the disproportionation of sulfur compounds of intermediate oxidation states (e.g., Finster 2008). In general, 10 to 50 percent of the global \ce{H2S} emission on Earth is as a metabolic byproduct, whereas over 70\% percent of natural \ce{SO2} emission is volcanic, although on modern Earth 90\% of atmospheric \ce{SO2} is anthropogenic (Seinfeld \& Pandis 2006). Understanding the atmospheric response to \ce{H2S} and \ce{SO2} emission will allow us to examine whether or not \ce{H2S} is a potential biosignature gas on a terrestrial exoplanet.

Previous investigation of sulfur photochemistry in the context of super Earth characterization has been very limited. Sulfur compounds are generally not considered in most models of terrestrial exoplanet atmospheres. Des Marais et al. (2002) describe the spectral features of certain molecules such as \ce{H2O}, \ce{N2O}, \ce{O3} and \ce{CH4} in terrestrial exoplanets. Miller-Ricci et al. (2009) present spectra of super Earths under equilibrium chemistry with photochemistry estimates and without considering sulfur compounds. Zahnle et al. (2009) investigate sulfur photochemistry in hot Jupiters and suggest that \ce{HS} and \ce{S2} can be generated photochemically from \ce{H2S}. Kaltenegger \& Sasselov (2010) study the effect of \ce{H2S} and \ce{SO2} on the terrestrial planetary spectra and suggest that 1-10 ppm \ce{SO2} has potentially detectable spectral features that could indicate active volcanism. Kaltenegger \& Sasselov (2010), however, do not link the  surface emission of sulfur gases to the mixing ratio of sulfur species via photochemistry involving sulfur compounds, or consider formation of aerosols. Domagal-Goldman et al. (2011) study the chemistry of organic sulfur compounds that are strongly linked to biology (e.g., \ce{CH3SH}) in anoxic \ce{N2} atmospheres and suggest that the most detectable feature of organic sulfur gases is their indirect photochemical product, ethane. Moreover, the fate of surface emission of sulfur compounds, e.g., \ce{H2S} and \ce{SO2}, is yet to be explored for \ce{H2}-dominated atmospheres. \ce{H2}-dominated atmospheres, like \ce{N2} or \ce{CO2} atmospheres with water vapor, could also maintain a habitable temperature at the surface through collision-induced absorption (Pierrehumbert \& Gaidos 2011; Wordsworth 2012). 

In all, it is still largely unknown whether or not \ce{H2S} and \ce{SO2} spectral features can be observed in the future in the atmosphere of an exoplanet and whether or not the emission rate of sulfur compounds on a terrestrial exoplanet can be inferred. In this paper we investigate the atmospheric chemistry resulting from \ce{H2S} and \ce{SO2} surface emission in atmospheres having very different oxidation states ranging from reducing to oxidizing. Here, ``emission" means the mass flux from the planetary surface to the atmosphere that may include volcanic sources and biogenic sources. We focus on terrestrial exoplanet atmospheres that include super Earths, since those planets hold the most interest on the path to finding and characterizing planets that might harbor life. In \S~2 we briefly describe our photochemistry model and radiative transfer model, with a focus of treatments of sulfur chemistry and aerosols. In \S~3 we describe the key chemistry pathways involving \ce{H2S} and \ce{SO2} in both reducing and oxidizing atmospheres. In \S~4 we present the main results on sulfur photochemistry in terrestrial exoplanet atmospheres and the spectral features of \ce{H2S}, \ce{SO2}, and photochemical aerosols. In \S~5 we discuss on whether or not \ce{H2S} can be a biosignature gas on a planet with atmospheric conditions different from Earth's. We present our conclusions in \S~6.

\section{Model}

\subsection{Photochemistry Model}

We have developed a comprehensive photochemistry model to investigate atmospheres of terrestrial exoplanets, and validated the model by simulating the atmospheric compositions of current Earth and Mars (see Paper I).  We now describe briefly the main features of the photochemistry model and the specifics for this work.

The purpose of the photochemistry model is to compute the steady-state chemical composition of an exoplanetary atmosphere. The system is described by a set of time-dependent continuity equations, one equation for species at each altitude. Each equation describes: chemical production; chemical loss; eddy diffusion and molecular diffusion (contributing to production or loss); sedimentation (for aerosols only); emission and dry deposition at the lower boundary; and diffusion-limited atmospheric escape for light species at the upper boundary. Starting from an arbitrary initial state, the system is numerically evolved to the steady state in which the number densities no longer change. Because the removal timescales of different species are very different, the implicit inverse Euler method is employed for the numerical time stepping. The generic model computes chemical and photochemical reactions among 111 molecules and aerosols made of O, H, N, C, S elements, and formation of sulfur (\ce{S8}) and sulfate (\ce{H2SO4}) aerosols. The numerical code is designed to have the capacity of treating both reducing and oxidizing atmospheres. For the chemical and photochemical reactions, we use the reaction rates data from both the NIST database (http://kinetics.nist.gov) and the JPL publication (Sander et al. 2011). We have also adopted relevant reaction rates from Kasting (1990), Yung \& DeMore (1999), and Moses et al. (2002). Sulfur polymerization reaction rates still lack consistent experimental measurements, and we adopt the reaction rates proposed by Kasting (1990) and those proposed by Moses et al. (2002). Both sets of sulfur polymerization reaction rates are speculative and they are widely discrepant, the effect of which will be discussed later in section 3.1. Ultraviolet and visible radiation in the atmosphere is computed by the $\delta$-Eddington 2-stream method, with molecular absorption, Rayleigh scattering and aerosol Mie scattering contributing to the opacity. For the stellar input spectrum we used the Air Mass Zero (AM0) reference spectrum produced by the American Society for Testing and Materials\footnote{http://rredc.nrel.gov/solar/spectra/am0/} for Sun-like stars, and used the simulated non-active M star spectrum from Allard et al. (1997) for quiet M stars. 

In this paper we use the photochemistry model to study the sulfur chemistry in atmospheres ranging from reducing to oxidizing on terrestrial exoplanets. We use \ce{H2} dominated atmospheres as the representative cases for reducing atmospheres, and we use \ce{N2}, and \ce{CO2} dominated atmospheres as the representative cases for oxidized atmospheres that could be both reducing and oxidizing. 
With the photochemistry model, we simulated the chemical composition of \ce{H2}, \ce{N2}, and \ce{CO2}-dominated atmospheres, with sulfur compounds emitted from the surface at various rates. Parameters of the atmospheric models are tabulated in Table \ref{AtmosPara} and details of water, carbon, and oxygen chemistry have been described in Paper I. We will focus on the sulfur chemistry and photochemistry in terrestrial exoplanet atmospheres in this paper.

\begin{table}[htdp]
\tiny
\caption{Basic parameters for the atmosphere models in this paper. The scenarios are \ce{H2}, \ce{N2}, \ce{CO2} dominated atmospheres on an Earth-sized atmosphere in the habitable zone of a Sun-like star. The planet has Earth-like volcanic emissions for hydrogen, oxygen, and carbon species, and various \ce{H2S} and \ce{SO2} emissions. Note that we do not consider any biotic contribution to the dry deposition velocities, and the volcanic carbon emission is not proportionally increased with sulfur emission. }
\begin{center}
\begin{tabular}{llll}
\hline\hline
Parameters & Reducing & Oxidized & Oxidized\\
\hline
Main component & 90\%\ce{H2}, 10\%\ce{N2}  & \ce{N2}  & 90\%\ce{CO2}, 10\%\ce{N2}  \\
Mean molecular mass & 4.6  & 28  & 42.4  \\
\hline
\multicolumn{4}{l}{\it Planetary parameters}\\
Stellar type & G2V & G2V & G2V \\
Semi-major axis & 1.6 AU & 1.0 AU & 1.3 AU \\
Mass & $M_{\earth}$ & $M_{\earth}$ &  $M_{\earth}$ \\
Radis & $R_{\earth}$ & $R_{\earth}$ & $R_{\earth}$ \\
\hline
\multicolumn{4}{l}{\it Temperature profile}\\
Surface temperature & 288 K & 288 K & 288 K \\
Surface pressure & $10^5$ Pa & $10^5$ Pa & $10^5$ Pa \\
Tropopause altitude & 120 km & 13.4 km & 8.7 km \\
Temperature above tropopause & 160 K & 200 K & 175 K \\
Maximum altitude &  440 km & 86 km & 51 km \\
\hline
\multicolumn{4}{l}{\it Eddy diffusion coefficient}\\
In the convective layer & $6.3\times10^5$ cm$^2$ s$^{-1}$ & $1.0\times10^5$ cm$^2$ s$^{-1}$ & $6.8\times10^4$ cm$^2$ s$^{-1}$ \\
Minimum & $2.5\times10^4$ cm$^2$ s$^{-1}$ & $3.9\times10^3$ cm$^2$ s$^{-1}$ &$2.7\times10^3$ cm$^2$ s$^{-1}$ \\
Altitude for the minimum & 107 km & 17.0 km & 11.6 km \\
Near the top of atmosphere & $7.1\times10^5$ cm$^2$ s$^{-1}$ & $1.1\times10^5$ cm$^2$ s$^{-1}$ & $7.6\times10^4$ cm$^2$ s$^{-1}$ \\
\hline
\multicolumn{4}{l}{\it Water and rainout}\\
Liquid water ocean & Yes & Yes & Yes \\
Water vapor boundary condition & $f(\ce{H2O})=0.01$ & $f(\ce{H2O})=0.01$ & $f(\ce{H2O})=0.01$ \\
Rainout rate\tablenotemark{a} & Earth-like & Earth-like & Earth-like \\
\hline
\multicolumn{4}{l}{\it Gas emission\tablenotemark{b} }\\
\ce{CO2} & $3\times10^{11}$ cm$^{-2}$ s$^{-1}$ & $3\times10^{11}$ cm$^{-2}$ s$^{-1}$ & N/A\\
\ce{H2} & N/A & $3\times10^{10}$ cm$^{-2}$ s$^{-1}$ & $3\times10^{10}$ cm$^{-2}$ s$^{-1}$ \\
\ce{CH4} & $3\times10^8$ cm$^{-2}$ s$^{-1}$ & $3\times10^8$ cm$^{-2}$  & $3\times10^8$ cm$^{-2}$ \\
\ce{SO2} & Vary &  Vary  &  Vary  \\
\ce{H2S} &  Vary &  Vary  &  Vary \\
\hline
\multicolumn{4}{l}{\it Dry deposition velocity\tablenotemark{c} }\\
\ce{H2} &\multicolumn{3}{l}{0} \\
\ce{CH4} &\multicolumn{3}{l}{0}\\
\ce{C2H6} &\multicolumn{3}{l}{ $1.0\times10^{-5}$ (Assumed) }\\
\ce{CO} &\multicolumn{3}{l}{$1.0\times10^{-8}$ cm s$^{-1}$ (Kharecha et al. 2005) } \\
\ce{CH2O} &\multicolumn{3}{l}{0.1 cm s$^{-1}$ (Wagner et al. 2002) } \\
\ce{CO2} &\multicolumn{3}{l}{$1.0\times10^{-4}$ cm s$^{-1}$ (Archer 2010) }\\
\ce{O2} &\multicolumn{3}{l}{0}\\
\ce{O3} & \multicolumn{3}{l}{0.4 cm s$^{-1}$ (Hauglustaine et al. 1994) } \\
\ce{H2O2} & \multicolumn{3}{l}{0.5 cm s$^{-1}$ (Hauglustaine et al. 1994) } \\
\ce{H2S} & \multicolumn{3}{l}{0.015 cm s$^{-1}$ (Sehmel 1980) }\\
\ce{SO2} & \multicolumn{3}{l}{1.0 cm s$^{-1}$ (Sehmel 1980) } \\
\ce{S8(A)} & \multicolumn{3}{l}{0.2 cm s$^{-1}$ (Sehmel 1980) } \\
\ce{H2SO4(A)} & \multicolumn{3}{l}{0.2 cm s$^{-1}$ (Sehmel 1980) }\\
\hline\hline
\end{tabular}
\tablenotetext{a}{Rainout rates for \ce{H2}, \ce{CO}, \ce{CH4}, \ce{C2H6}, and \ce{O2} are generally assumed to be zero to simulate an ocean surface saturated with these gases on an abiotic exoplanet. }
\tablenotetext{b}{The volcanic gas emission rates from the planetary surface are assigned for each model scenario. \ce{H2O} emission is not explicitly considered because the surface has a large water reservoir, i.e., an ocean.}
\tablenotetext{c}{We here list the dry deposition velocities (with references) for emitted gases and their major photochemical byproducts, and dry deposition velocities that are important for the mass and redox balance of the atmosphere. Dry deposition velocities are assumed to be identical for the three scenarios. \ce{C2H6} dry deposition velocity is assumed to take into account the loss of carbon due to organic haze formation and deposition. The \ce{CO2} dry deposition velocity is assumed such that the steady-state mixing ratio of \ce{CO2} in \ce{H2} and \ce{N2} atmospheres is in the order of 100 ppm. }
\end{center}
 \label{AtmosPara}
\end{table}

We consider in our atmospheric chemistry models the O, H, and S bearing species and a subset of C bearing species. The gaseous molecules considered in this paper are \ce{H}, \ce{H2}, \ce{O}, \ce{O(^1D)}, \ce{O2}, \ce{O3}, \ce{OH}, \ce{HO2}, \ce{H2O}, \ce{H2O2}, \ce{CO2}, \ce{CO}, \ce{CH2O}, \ce{CHO}, \ce{C}, \ce{CH}, \ce{CH2}, \ce{^1CH2}, \ce{CH3}, \ce{CH4}, \ce{CH3O}, \ce{CH4O}, \ce{CHO2}, \ce{CH2O2}, \ce{CH3O2}, \ce{CH4O2}, \ce{C2}, \ce{C2H}, \ce{C2H2}, \ce{C2H3}, \ce{C2H4}, \ce{C2H5}, \ce{C2H6}, \ce{C2HO}, \ce{C2H2O}, \ce{C2H3O}, \ce{C2H4O}, \ce{C2H5O}, \ce{S}, \ce{S2}, \ce{S3}, \ce{S4}, \ce{SO}, \ce{SO2}, \ce{^1SO2}, \ce{^3SO2}, \ce{SO3}, \ce{H2S}, \ce{HS}, \ce{HSO}, \ce{HSO2}, \ce{HSO3}, \ce{H2SO4}, and \ce{S8}, and the aerosols considered are \ce{S8} aerosols and \ce{H2SO4} aerosols. This set of species is comprised of common H, O, and C bearing species and photochemical products of \ce{H2S} and \ce{SO2} emission. We assume a constant \ce{H2O} relative humidity at the surface of 60\% to mimic the supply of water vapor from a liquid water ocean. To reduce the stiffness of the system and improve the numerical stability, ``fast" species with relatively short chemical loss timescales are computed directly from the photochemistry equilibrium. We consider in this work \ce{O(^1D)}, \ce{^1CH2}, \ce{C2H}, \ce{^1SO2}, and \ce{^3SO2} as fast varying species. As such, the photochemistry model rigorously finds steady-state composition of the atmosphere starting with initial compositions without any sulfur compounds. Once the model converged to the steady state, we checked explicitly the mass conservation of O, H, C, N, S atoms and verified the choice of fast species to have been appropriate. We have required all models to balance mass flux within $10^{-3}$ for convergence, and typically our models balance mass flux to $10^{-6}$. We have also explicitly checked the redox (i.e., hydrogen budget) balance (see our definition of the redox number, flux, and balance in Paper I), and required the models to balance redox flux to $10^{-3}$.

One of the most significant controlling factors that determine the steady-state composition of atmospheres on terrestrial exoplanets is the dry deposition velocities of emitted gases and their major photochemical byproducts. Of particular importance in this paper is the dry deposition velocities of \ce{H2S} and \ce{SO2}, which could vary by orders of magnitude (see Table \ref{AtmosPara} for the fiducial values of key dry deposition velocities).
The dry deposition velocities depend on the properties of the lower atmosphere and the surface. For example, in a model of the early cold Martian atmosphere, the deposition velocities are assumed to be reduced by an artificial factors of up to 1000 compared with those in warm current Earth (e.g., Tian et al. 2010), in order to account for less efficient deposition at a lower temperature. For another example, in modeling the early Martian atmosphere, because the putative Mars ocean is believed to be saturated with dissolved \ce{SO2} and other sulfur species, \ce{SO2} deposition is assumed to be balanced by the an equivalent return flux from the ocean (e.g. Halevy et al. 2007; Tian et al. 2010) and then $V_{\rm DEP}$ of \ce{SO2} is assumed to be zero. In this paper, we explore the effect of varying the dry deposition velocities of \ce{H2S} and \ce{SO2}, which can be scaled down by a factor of 100 if the surface is saturated with sulfide or sulfite. 

Another important factor that strongly influences the atmospheric sulfur chemistry is the formation and sedimentation of aerosols. Photochemically produced \ce{H2SO4} and \ce{S8} may condense to form aerosols if their concentrations exceed their vapor saturation concentrations. The parameterization of condensation and sedimentation of aerosols in our photochemistry model is described in detail in Paper I. 
Saturation vapor pressure of \ce{H2SO4} is taken as recommended by Seinfeld \& Pandis (2006) for atmospheric modeling, with a validity temperature range of 150 - 360 K. 
\ce{S8} is the stable form of elemental sulfur because the \ce{S8} molecule has a crown-shape ring structure that puts least strain on the S-S bond among sulfur allotropes and the crown structure allows for considerable cross-ring interaction between nonbonded atoms (Meyer 1976).
The saturation pressure of \ce{S8} is then taken as the total sulfur saturation pressure against liquid sulfur at $T>392$ K and solid (monoclinic) sulfur at $T<392$ K tabulated by Lyons (2008). 
We consider the average aerosol particle diameter, a key parameter that determines the aerosols' dynamical and optical properties, to be a free parameter. On Earth, the ambient aerosol size distribution is dominated by several modes corresponding to different sources. The ``condensation submode", formed from vapor condensation and coagulation, has an average diameter of $\sim0.4$ $\mu$m (Seinfeld \& Pandis, 2006). On Venus, the Mode 1 particles with an an average diameter of $\sim1.0$ $\mu$m dominate the upper cloud (e.g., Carlson et al. 1993). On Titan, the photochemical aerosols in the stratosphere have mean diameters in the range of 0.1 - 1 $\mu$m (Rages et al. 1983). We treat the particle diameter a free parameter, and explore the effects of varying the particle diameter from 0.1 to 10 $\mu$m. The dry deposition velocity of aerosols is assumed to be 0.2 cm s$^{-1}$, a sensible deposition velocity of particles having diameters between 0.1 and 1 $\mu$m (Sehmel 1980; Seinfeld \& Pandis 2006). 

Before leaving this section we provide the physical rationale of specifying temperature-pressure profiles and eddy diffusion coefficients for our atmospheric photochemistry models.

We modeled the atmospheric composition from the 1-bar pressure level up to the altitudes of about 10 scale heights. We chose appropriate vertical resolution for each scenario so that there are 4 layers per scale height. We adopted a temperature profile for our atmosphere models, without considering feedback on temperature of the atmospheric composition. The surface temperature is assumed to be 288 K. The semi-major axis of a terrestrial exoplanet around a Sun-like star implied by this surface temperature is 1.6 AU, 1.0 AU, and 1.3 AU, for \ce{H2}, \ce{N2}, and \ce{CO2} dominated atmosphere, estimated based on a similar procedure as Kasting et al. (1993). The temperature profiles are then assume to follow appropriate dry adiabatic lapse rate (i.e., the convective layer) until 160 K (\ce{H2} atmosphere), 200 K (\ce{N2} atmosphere), and 175K (\ce{CO2} atmosphere) and to be constant above (i.e., the radiative layer). The adopted temperature profiles are consistent with significant greenhouse effects in the convective layer and no additional heating above the convective layer for habitable exoplanets. We did not consider the climate feedback of \ce{SO2} or sulfur aerosols in the atmosphere, which could be important to determine the surface temperature (e.g. Halevy et al. 2007; Tian et al. 2010). While not ideal, these temperature profiles yield the same results discussed in the below as temperature profiles varied by several tens of K. The precise temperature-pressure structure of the atmosphere is less important than photochemistry for the investigation of sulfur chemistry because the most important photolysis and chemical reactions are not significantly affected by minor deviations in the temperature profile. We found that variation of temperature profiles by a few tens of K has minor impact on the atmospheric composition.

Vertical transport of gases in the atmosphere is parameterized by eddy diffusion, and the coefficients are assumed to be those of Earth's atmosphere scaled by the atmospheric scale height to account for \ce{H2}, \ce{N2}, and \ce{CO2} being the dominant species. The current Earth's eddy diffusion coefficient profile has been derived from mixing ratio profiles of several long-lived gases (Massie \& Hunten 1981; also shown in Figure 1 of Paper I). We use the empirical eddy diffusion coefficient profile for current Earth as a template, and scale the coefficient inversely with the mean molecular mass for \ce{H2}, \ce{N2}, and \ce{CO2} dominated atmospheres. 
The justification for such scaling is that the eddy diffusion coefficient is proportional to the mixing length which is in turn a portion of the atmospheric scale height (e.g., Smith 1998).  The pressure surface to pressure surface projection also ensures that the eddy diffusion coefficient profile features a eddy diffusion minimum near the tropopause for atmospheres with different mean molecular masses. Our approach to parameterize vertical transport is of course an approximation. An accurate representation of vertical transport would likely involve circulation on the global scale (e.g., Holton 1986). We explore the effect of eddy diffusion coefficients ranging one or two orders of magnitude from the nominal value in a sensitivity study (see section 4.1).

\subsection{Radiative Transfer Model}

We compute synthetic spectra of the modeled exoplanet's atmospheric transmission, reflection and thermal emission with a line-by-line radiative transfer code (Seager \& Sasselov, 2000; Seager et al. 2000; Miller-Ricci et al. 2009; Madhusudhan \& Seager 2009). Opacities are based on molecular absorption with cross sections computed based from the HITRAN 2008 database (Rothman et al. 2009), molecular collision-induced absorption when necessary (e.g., Borysow 2002), Rayleigh scattering, and aerosol extinction are computed based on the Mie theory (e.g., Van de Hulst 1981). The transmission is computed for each wavelength by integrating the optical depth along the limb path, as outlined in Seager \& Sasselov (2000). The reflected stellar light and the planetary thermal emission are computed by the $\delta$-Eddington 2-stream method (Toon et al. 1989). We used the refractive index of \ce{S8} aerosols from Tian et al. (2010) for the UV and visible wavelengths and from Sasson et al. (1985) for infrared (IR) wavelengths. We used the refractive index of \ce{H2SO4} aerosols (assumed to be the same as 75\% sulfuric acid solution) from Palmer \& William (1975) for  UV to IR wavelengths, and Jones (1976) for far IR wavelengths.

\begin{figure}[h]
\begin{center}
 \includegraphics[width=0.5\textwidth]{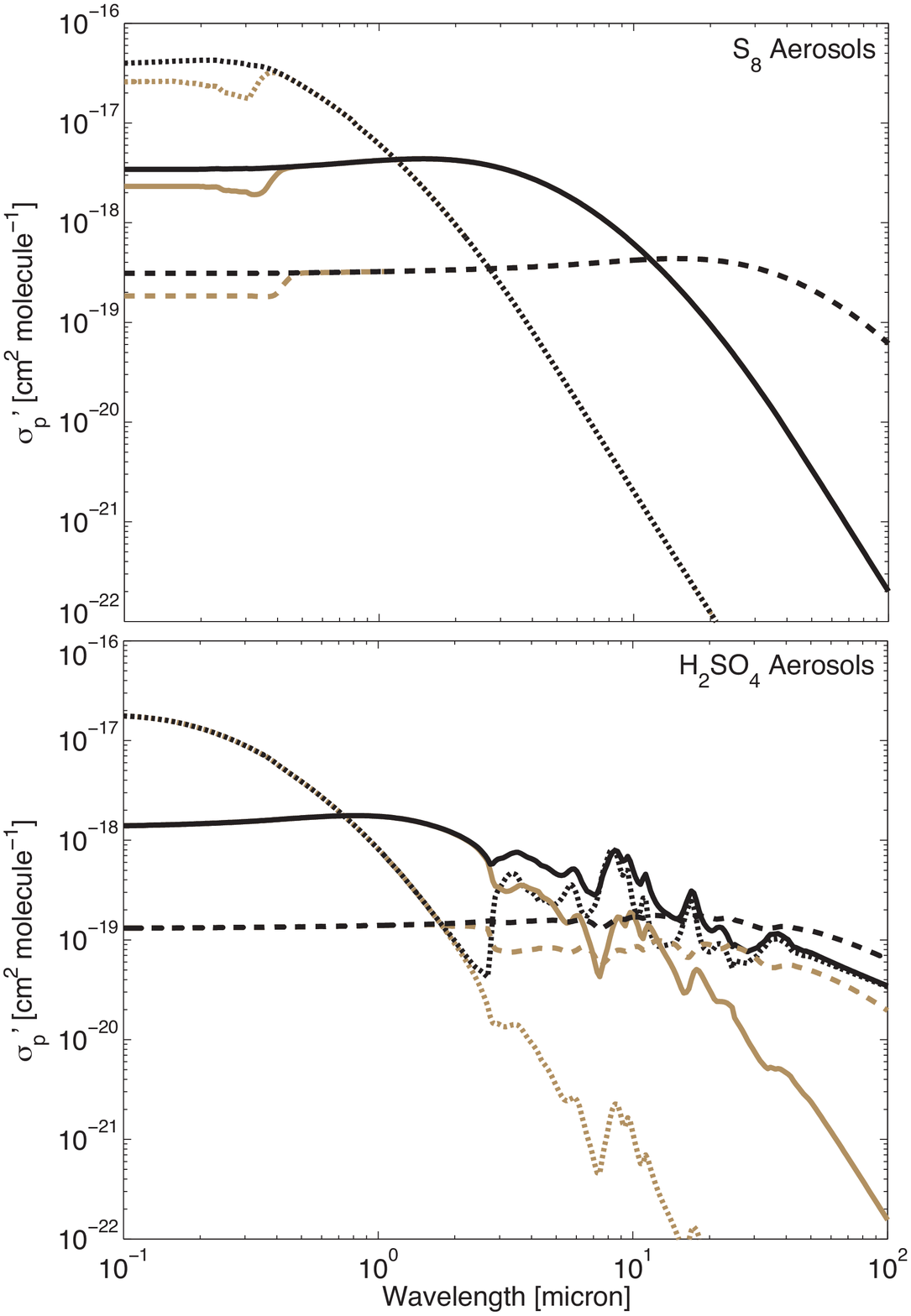}
 \caption{
 Extinction cross sections (black lines) and scattering cross sections (orange lines) of \ce{H2SO4} and \ce{S8} per molecule in the condensed phase.
 The dotted, solid, and dashed lines are cross sections for the mean particle diameter of 0.1, 1, 10 $\mu$m, respectively. 
 The size distribution of aerosol particles is assumed to be lognormal with a dispersion $\sigma=2$.
 }
 \label{Cross_compare}
  \end{center}
\end{figure}

The particle size distribution of aerosols controls their optical properties. We adopt the lognormal distribution as
\begin{equation}
\frac{dN}{dD} = \frac{N_t}{\sqrt{2\pi}D\ln\sigma}\exp\bigg[-\frac{(\ln D - \ln D_0)^2}{2\ln^2\sigma}\bigg],
\end{equation}
where $dN$ is the number of particles per volume in the diameter bin $dD$, $N_t$ is the total number density of particles, $D_0$ is the median diameter of the particles, and $\sigma$ is the particle size dispersion (defined as the ratio of the diameter below which 84.1\% of the particles lie to the median diameter). The lognormal distribution is a reasonable assumption because it provides a good fit to the particle size distribution measured in Earth's atmosphere (e.g., Seinfeld \& Pandis 2006), and a sensible particle size dispersion parameter for photochemically produced aerosols is in the range of $1.5\sim2.0$ (Seinfeld \& Pandis 2006). 	What is important for the radiative transfer model and the photochemistry model is the surface area mean diameter $D_S$ and the volume mean diameter $D_V$, respectively. The mean diameters are related to the median diameter as
\begin{equation}
D_S = D_0 \exp(\ln^2\sigma),
\end{equation}
\begin{equation}
D_V = D_0 \exp\bigg(\frac{3}{2}\ln^2\sigma\bigg).
\end{equation}
We use the surface area mean diameter $D_S$ (referred to as ``mean diameter" in the following) as the free parameter for specifying a particle size distribution, as it is relevant to the radiative properties of the particle population. The volume mean diameter  $D_V$ is useful in the conversion from mass concentration of the condensed phase (computed in the photochemistry model) to the number of aerosol particles for radiative transfer computation. The extinction cross sections of \ce{H2SO4} and \ce{S8} molecules in aerosols for various mean particular diameters are shown in Figure \ref{Cross_compare}.

Elemental sulfur aerosols and sulfuric acid aerosols have different optical properties at the visible and infrared (IR) wavelengths. In the visible, \ce{S8} aerosols have a larger cross section than \ce{H2SO4} aerosols (Figure \ref{Cross_compare}). For wavelengths less than 400 nm \ce{S8} aerosols are both reflective and absorptive. In the infrared, the cross section of \ce{S8} aerosols drops significantly with increasing wavelength unless the mean diameter is in the order of 10 $\mu$m. In contrast, \ce{H2SO4} aerosols have an enhancement of absorption at the MIR wavelengths (5-10 $\mu$m) for all particle sizes (Figure \ref{Cross_compare}). 


\section{Sulfur Chemistry in Reducing and Oxidizing Atmospheres}

We now briefly describe the most important processes of sulfur chemistry that occur in reducing and oxidizing atmospheres on rocky exoplanets. The primary sulfur emission from the planetary surface would be \ce{SO2} and \ce{H2S}; they are either deposited back to the surface via dry or wet deposition, or converted into other forms of sulfur compounds in the atmosphere by photochemical reactions. One of the main purposes of this paper is to study the fate of sulfur gases emitted from the surface and their possible photochemical byproducts in the atmosphere.

The fate of sulfur gases emitted from the surface is mainly controlled by the redox power of the atmosphere -  the ability to reduce or oxidize a gas in the atmosphere. The redox power, in turn, is controlled by both the main component in the atmosphere (e.g., \ce{H2}, \ce{N2}, and \ce{CO2}) and the surface emission and deposition of trace gases (i.e., \ce{H2}, \ce{CH4}, and \ce{H2S}), as shown in Table \ref{AtmosRedox}. In the extreme cases of the atmospheric redox state, i.e., the \ce{H2}-dominated atmospheres and the \ce{O2}-rich atmospheres, the atmospheric redox power is surely reducing or oxidizing, regardless of the nature of surface emission or deposition. However, for an intermediate redox state, the atmosphere would be composed of redox-neutral species like \ce{N2} and \ce{CO2}, and the redox power of the atmosphere can be mainly controlled by the emission and the deposition fluxes of trace gases from the surface. The higher the emission of reducing gases is, the more reducing the atmosphere becomes.

We already know that sulfur gas emissions are effectively oxidized into sulfate, the most oxidized form of sulfur, in the oxic atmospheres such as the Earth's (e.g., Seinfeld \& Pandis 2006). In anoxic atmospheres, which include the reduced atmospheres and the oxidized atmospheres, previous studies have shown that both elemental sulfur and sulfate could be formed (Kasting 1990; Pavlov \& Kasting 2002; Zahnle et al. 2006; Hu et al. 2012). For this paper, we use \ce{H2} dominated atmospheres as the representative cases for reducing atmospheres, and we use \ce{N2}, and \ce{CO2} dominated atmospheres as the representative cases for oxidized atmospheres that could be both reducing and oxidizing. We now describe the key sulfur chemistry processes in these atmospheres as the follows.

\begin{table}[htdp]
\tiny
\caption{Redox power of atmospheres on rocky exoplanets.}
\begin{center}
\begin{tabular}{llllll}
\hline\hline
Type & Main Component & Redox Power & Main Reactive Species & Solar-System analogs & Note\\
\hline\hline
Reduced & \ce{H2}, \ce{CO} & Reducing & \ce{H} & None & \\
\hline
Oxidized & \ce{N2}, \ce{CO2} & Weakly reducing & \ce{H}, \ce{OH}, \ce{O} & None & The redox power is mainly controlled by the \\
                  &                                       & Weakly oxidizing & \ce{H}, \ce{OH}, \ce{O} & Mars, Venus & surface emission of trace gases (\ce{H2}, \ce{CH4}, \ce{H2S}).\\
\hline
Oxic & \ce{O2} & Highly oxidizing & \ce{OH}, \ce{O} & Earth & \\
\hline
\hline
\end{tabular}
\end{center}
\label{AtmosRedox}
\end{table}

\subsection{Reducing \ce{H2}-Dominated Atmospheres}

\begin{figure}[h]
\begin{center}
 \includegraphics[width=0.5\textwidth]{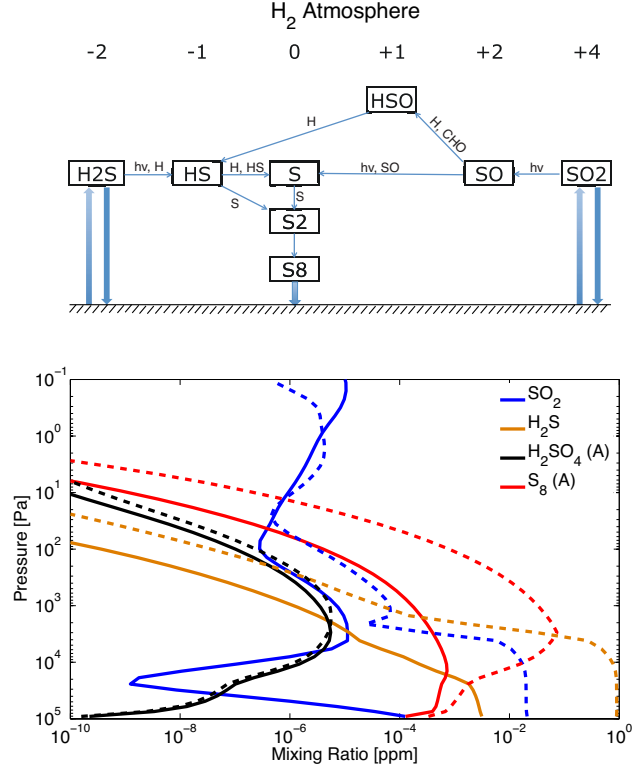}
 \caption{
Formation of elemental sulfur aerosols in reducing \ce{H2}-dominated atmospheres on an Earth-sized rocky planet in the habitable zone of a Sun-like star. The upper panel schematically illustrates the chemical pathways from the primary sulfur emission (i.e., \ce{H2S} and \ce{SO2}) to elemental sulfur, in which sulfur compounds are located according to their oxidation states labeled on the top of the figure. The thin arrows show the major chemical pathways in the atmosphere, and the thick arrows show the major surface-atmosphere interactions.
The lower panel shows the results of photochemistry simulations, with total surface sulfur emission of $10^{10}$ (solid lines) and $10^{12}$ (dashed lines) cm$^{-2}$ s$^{-1}$, i.e, 3 and 300 times higher than the Earth's volcanic sulfur emission rate. The \ce{H2S}/\ce{SO2} ratio in the sulfur emission is 0.5 and the particle mean diameter is 0.1 $\mu$m. Other model parameters are tabulated in Table \ref{AtmosPara}.
UV photons and atomic hydrogen effectively convert \ce{H2S} and \ce{SO2} into elemental sulfur, and elemental sulfur aerosols shield UV photons so that \ce{H2S} and \ce{SO2} may accumulate below the aerosol layer if the sulfur emission is more than 300 times higher than the Earth's volcanic emission rate. 
 }
 \label{H2_Chem}
  \end{center}
\end{figure}

Both \ce{H2S} and \ce{SO2} emitted from the surface  are efficiently converted into elemental sulfur in reducing \ce{H2} atmospheres. The major chemical pathways for sulfur compounds in the \ce{H2}-dominated atmosphere and the results of photochemistry model simulations are shown in Figure \ref{H2_Chem}.  Atomic hydrogen produced from photodissociation of water vapor and \ce{H2S} itself is the key reactive species that converts \ce{H2S} and \ce{SO2} into elemental sulfur. The primary chemical loss for \ce{H2S} in the atmosphere is via
\reaction{H2S + h$\nu$ -> HS + H ,\label{H2SLoss1}}
and
\reaction{H2S + H -> HS + H2 .\label{H2SLoss2}}
The \ce{HS} produced can then react with \ce{H} again or with itself to produce elemental sulfur. \ce{HS} can also react with \ce{S} to produce \ce{S2}. The primary chemical loss for \ce{SO2} in the atmosphere is photodissociation that produces \ce{SO}. \ce{SO} can be either photodissociated to elemental sulfur, or be further reduced to \ce{HS} via \ce{HSO} by H or \ce{CHO} and then converted to elemental sulfur (see Figure \ref{H2_Chem}). 

The \ce{S} and \ce{S2} molecules produced in the atmosphere will polymerize to form \ce{S8}, and \ce{S8} will condense to form aerosols if it is saturated in the atmosphere. Due to its ring structure, \ce{S8} is stable against photodissociation. \ce{S8} is a strong UV absorber (Kasting 1990). Therefore, \ce{S8} aerosols, if produced in the atmosphere, can effectively shield UV photons so that \ce{H2S} and \ce{SO2} may accumulate beneath the aerosol layer (see the case for a sulfur emission rate 300 times higher than Earth's current volcanic sulfur emission rate shown in Figure \ref{H2_Chem}).

The primary source of atomic hydrogen is the photolysis of \ce{H2O}, which occurs above the altitudes of $\sim10^3$ Pa pressure level. The atomic hydrogen can be then transported by eddy diffusion to the pressure level of $\sim10^4$ Pa to facilitate the removal of \ce{H2S} and \ce{SO2} and the production of elemental sulfur. Additional numerical simulations show that an increase of the eddy diffusion coefficient by one order of magnitude can increase the yield of elemental sulfur by about 20\%, because the transport of atomic hydrogen becomes more efficient. A secondary source of atomic hydrogen is photolysis of \ce{H2S} (reaction \ref{H2SLoss1}). This secondary source for atomic hydrogen is particularly important when the host star is a quiet M dwarf, because a quiet M dwarf produces few photons that could dissociate water\footnote{Water is principally dissociated by photons in the 150 - 200 nm wavelength range, whereas \ce{H2S} is principally dissociated by photons in the 200 - 260 nm wavelength range.}. For planets around quiet M dwarfs the photolysis of \ce{H2S} could be the main source of atomic hydrogen in their atmospheres. Additional numerical simulations show that in the habitable zone of a quiet M dwarf having an effective temperature of 3100 K, \ce{H2S} photolysis alone can produce enough atomic hydrogen to drive the formation of elemental sulfur in the atmosphere.

We here comment on the uncertainty of photochemistry models regarding the yield of \ce{S8}. In our model, we have assumed polymerization of elemental sulfur proceeds via
\reaction{S + S -> S2 \ , \label{SP0}}
\reaction{S + S2 -> S3 \ , \label{SP1}}
\reaction{S + S3 -> S4 \ ,\label{SP2}}
\reaction{S2 + S2 -> S4 \ ,\label{SP3}}
\reaction{S4 + S4 -> S8 \ .\label{SP4}}
We also include photodissociation for \ce{S2}, \ce{S3}, and \ce{S4}. However, the reaction rates of sulfur polymerization (reactions \ref{SP0} - \ref{SP4}) have not been well established by laboratory studies, and previous authors have adopted different rate constants for these reactions. In particular, Kasting (1990) and Pavlov \& Kasting (2002) have used 3-order-of-magnitude lower rates for reactions (\ref{SP0} - \ref{SP1}) and 1-order-of-magnitude lower rates for reactions (\ref{SP2} - \ref{SP4}), compared with Moses et al. (2002). In this work, we have adopted the reaction rates of Moses et al. (2002) for elemental sulfur reactions. Our sensitivity tests show that adopting the reaction rates of Kasting (1990) would result in about 3 to 10 times less \ce{S8}. We have chosen a higher sulfur polymerization rates for nominal models because: (1) the chemical pathways of reactions (\ref{SP0} - \ref{SP4}) are probably not complete and there may be other pathways to form \ce{S8}; (2) \ce{S2}, \ce{S3}, and \ce{S4} may condense as suggested by Lyons (2008) and the polymerization may still proceed in the condensed phase to \ce{S8}. Experimental studies are encouraged to settle this important uncertainty.

\subsection{Oxidized \ce{N2} and \ce{CO2}-Dominated Atmospheres }

\begin{figure}[h]
\begin{center}
 \includegraphics[width=0.5\textwidth]{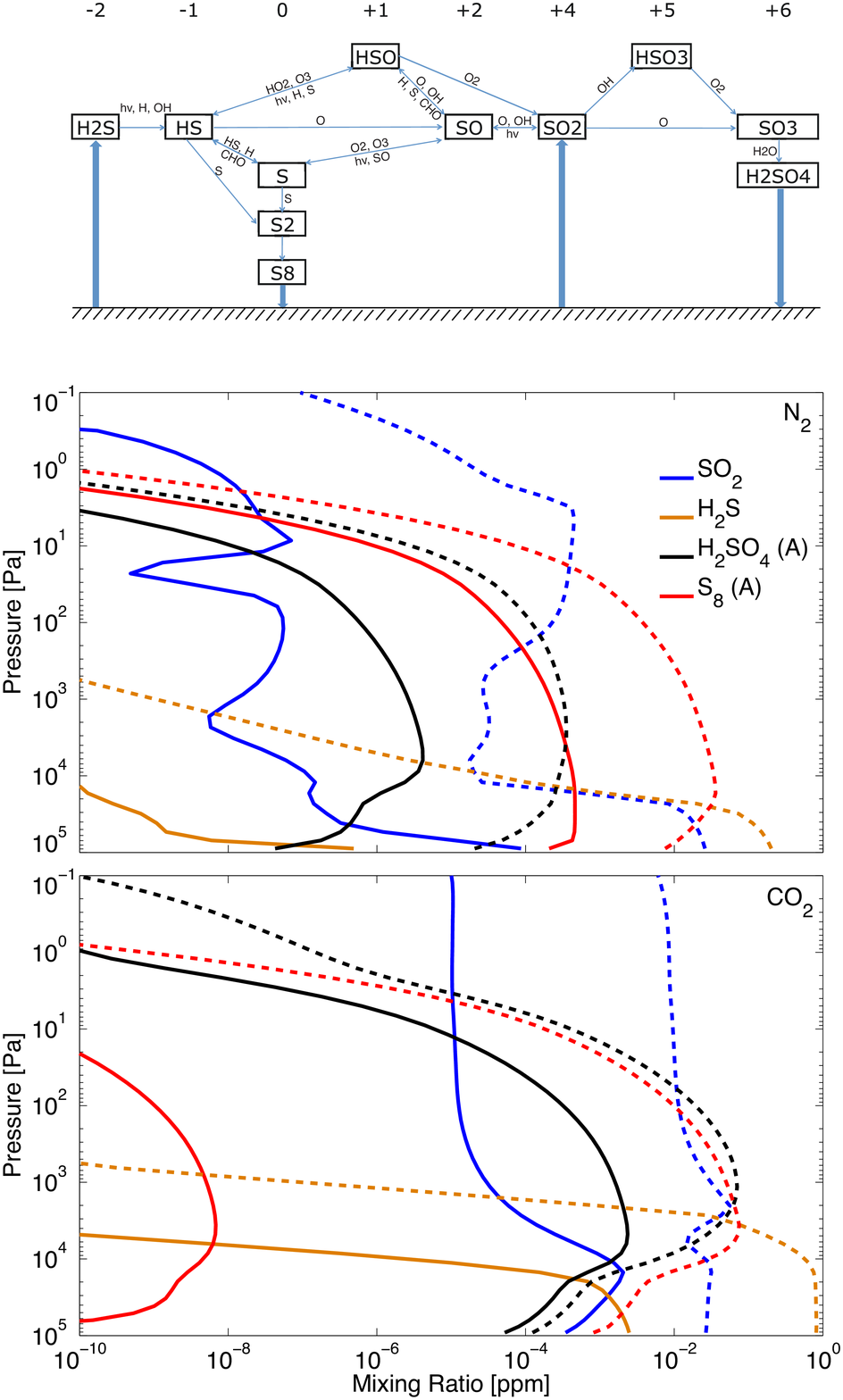}
 \caption{
Formation of sulfuric acid aerosols in oxidized \ce{N2} and \ce{CO2} dominated atmospheres on an Earth-sized rocky planet in the habitable zone of a Sun-like star.
Similar to Figure \ref{H2_Chem}, the upper panel schematically illustrates the chemical pathways from the primary sulfur emissions (i.e., \ce{H2S} and \ce{SO2}) to elemental sulfur and sulfuric acid. For double arrows the label above the arrow indicates the oxidizing agents, and the label below the arrow indicates the reducing agents.
The lower two panel shows the results of photochemistry simulations for \ce{N2} and \ce{CO2} dominated atmospheres, respectively. The total surface sulfur emission is $10^{10}$ (solid lines) and $10^{12}$ (dashed lines) cm$^{-2}$ s$^{-1}$, i.e, 3 and 300 times higher than the Earth's volcanic sulfur emission rate. The \ce{H2S}/\ce{SO2} ratio in the sulfur emission is 0.5 and the particle mean diameter is 0.1 $\mu$m. Other model parameters are tabulated in Table \ref{AtmosPara}.
Both elemental sulfur aerosols and sulfuric acid aerosols are formed in the oxidized and anoxic atmospheres. The origins of the principle reducing agents (\ce{H} and \ce{CHO}) and the principle oxidizing agents (\ce{OH}, \ce{O} and \ce{O2}) is photodissociation of \ce{H2O} and \ce{CO2}. The apparent depletion of \ce{SO2} at the pressure level of 10 - 100 Pa in the \ce{N2} atmosphere (the blue solid line in the middle panel) is due to the production of atomic hydrogen by methane photodissociation at this pressure level.
 }
 \label{N2_Chem}
  \end{center}
\end{figure}

\ce{H2S} and \ce{SO2} gases emitted from the surface can be converted into both elemental sulfur (\ce{S8}) and sulfuric acid (\ce{H2SO4}) in the oxidized (but anoxic) atmospheres such as \ce{N2} and \ce{CO2} dominated atmospheres. The major chemical pathways that lead to formation of both elemental sulfur and sulfuric acid, and the results of photochemistry model simulations are shown in Figure \ref{N2_Chem}. The production of elemental sulfur aerosols involves UV photons and atomic hydrogen, as does in reducing \ce{H2} atmospheres; whereas the production of sulfuric acid requires oxidizing species, notably \ce{OH} and \ce{O2}. These reactive species, either reducing or oxidizing, are produced from photolysis of water and \ce{CO2}. In particular, the source of \ce{OH}, responsible for converting \ce{SO2} to sulfuric acid in the atmosphere, is the photodissociation of \ce{H2O}. As a result, the amount of UV photons that are capable of dissociating water controls the yield of \ce{H2SO4}. For example in the habitable zone of a quiet M dwarf the yield of \ce{H2SO4} is much reduced compared with solar-like stars by at least one order of magnitude. 

The photochemically produced \ce{S8} and \ce{H2SO4} may condense to form aerosols in the atmosphere if saturated. As a result, aerosols in the atmospheres provide a UV shield that enables the accumulation of \ce{H2S} and \ce{SO2} beneath the layer of aerosols. In particular for an Earth-sized planet in the habitable zone of a Sun-like star, when the surface emission rate is more than two-orders-of-magnitude higher than the current Earth's volcanic sulfur emission rate, photochemical aerosols in the atmosphere lead to substantial UV shielding for accumulation of \ce{H2S} and \ce{SO2}, as shown in Figure \ref{N2_Chem}. We find that only when sulfur emission is highly elevated with respect to current Earth could \ce{H2S} or \ce{SO2} accumulate to the order of parts per million mixing ratio in the \ce{N2} and \ce{CO2} atmospheres.

\begin{figure}[h]
\begin{center}
 \includegraphics[width=0.5\textwidth]{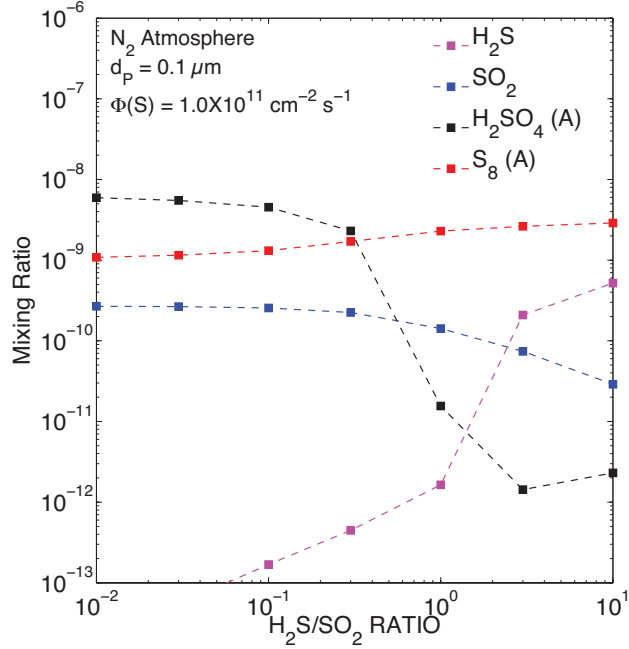}
 \caption{Correlation between the aerosol composition and the composition of sulfur emissions in the weakly oxidizing
 \ce{N2} atmosphere on an Earth-sized rocky planet in the habitable zone of a Sun-like star. The total surface emission rate is $10^{11}$ cm$^{-2}$ s$^{-1}$, or 30 times the Earth's volcanic sulfur emission rate, and the particle mean diameter is 0.1 $\mu$m. Other model parameters are tabulated in Table \ref{AtmosPara}.
As a larger fraction of surface sulfur emission is in the form of \ce{H2S}, the amount of \ce{S8} aerosols in the atmosphere increases, and the amount of \ce{H2SO4} aerosols in the atmosphere decreases dramatically.
 }
 \label{N2_RATIO}
  \end{center}
\end{figure}

The relative yield between elemental sulfur and sulfuric acid is controlled by the redox power of the atmosphere. In general, more sulfuric acid aerosols and less elemental sulfur aerosols are anticipated in a more oxidizing atmosphere. The redox power of the atmosphere, in turn, is determined by both the main constituent and the reducing gas emission. \ce{CO2} dominated atmospheres are mosre oxidizing than \ce{N2} dominated atmospheres as photodissociation of \ce{CO2} leads to atomic oxygen. Therefore the primary sulfur emission is more likely to be converted to sulfuric acid in \ce{CO2} dominated atmospheres than in \ce{N2} dominated atmospheres (see Figure \ref{N2_Chem}). Surface emission of reducing gases, including \ce{H2}, \ce{CH4}, \ce{CO}, and \ce{H2S}, alters the redox budget of the atmosphere and therefore increases the relative yield of elemental sulfur versus sulfuric acid aerosols (e.g., Zahnle et al 2006). As shown in Figure \ref{N2_Chem}, when the sulfur emission rate increases, both \ce{N2} and \ce{CO2} atmospheres become more and more reducing (because \ce{H2S} is reducing), which results in a dramatic increase of elemental sulfur production in the atmosphere. Furthermore, the \ce{H2S}/\ce{SO2} ratio in the sulfur emission affects its contribution to the redox power of the atmosphere and then the relative abundances of the two types of aerosols in the atmosphere significantly. As a result of the increase in the \ce{H2S}/\ce{SO2} emission ratio, the amount of \ce{S8} aerosol in the atmosphere increases, and the amount of \ce{H2SO4} in the atmosphere decreases dramatically (see Figure \ref{N2_RATIO}). For an Earth-like planet having an \ce{N2} atmosphere, if the \ce{H2S}/\ce{SO2} emission ratio is less than 0.1 (as is the case for current Earth; Holland 2002), the dominant type of aerosols in the atmosphere is sulfate; whereas elemental sulfur aerosols become the dominant type if the \ce{H2S}/\ce{SO2} emission ratio is larger than 1.

\section{Results}

\label{Result}

\subsection{Optically Thick Aerosols from Sulfur Emission}

\begin{figure}[h]
\begin{center}
 \includegraphics[width=1.0\textwidth]{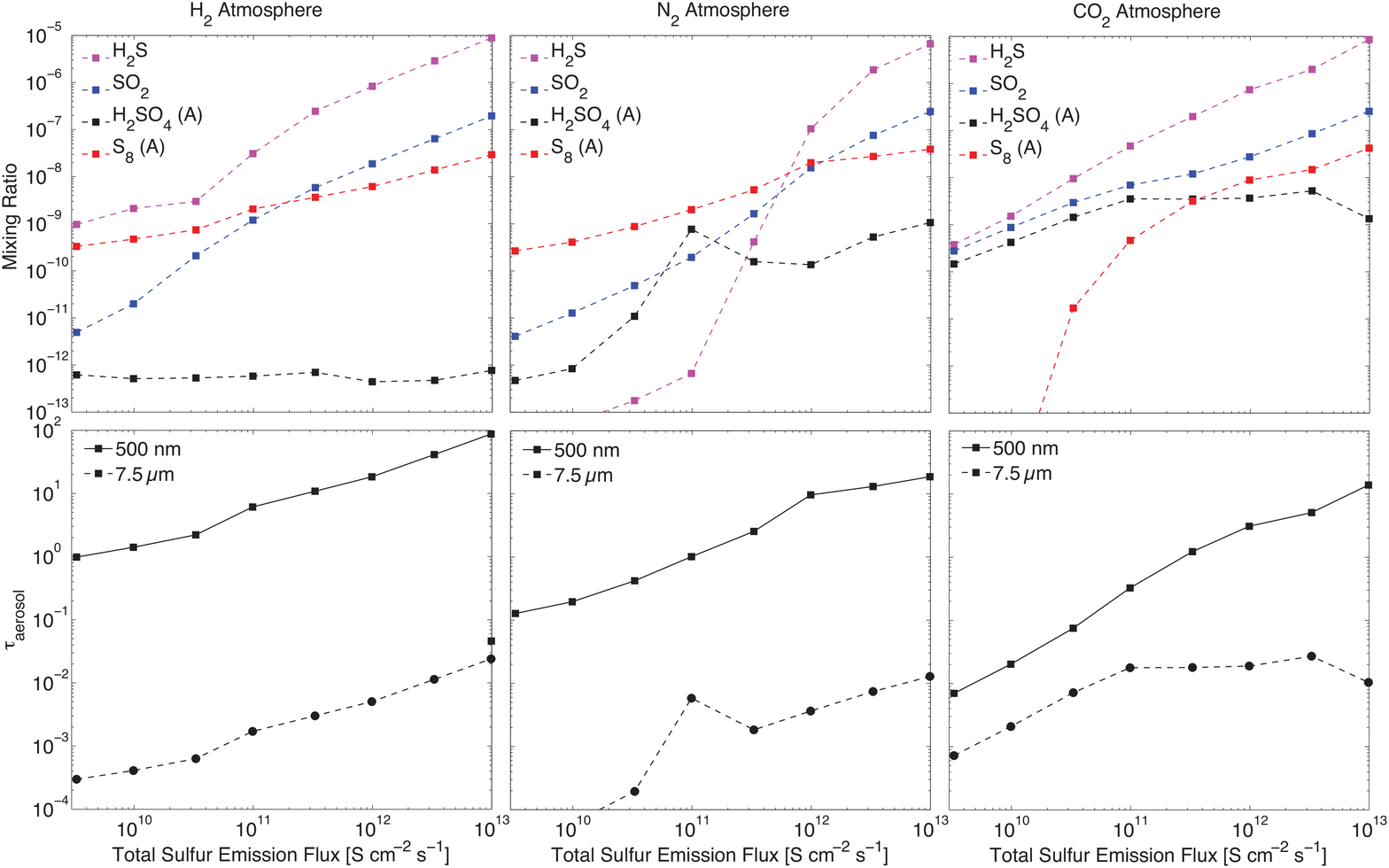}
 \caption{
 The relationship between the aerosol mixing ratios and aerosol opacities and the total sulfur emission rate.
 Column-integrated aerosol opacities at the 1-bar pressure level including both elemental sulfur aerosols and sulfuric acid aerosols at 500 nm (solid lines) and 7.5 $\mu$m (dashed lines) are shown in the lower panel. The planet is an Earth-sized rocky planet orbiting a Sun-like star, with reducing (\ce{H2}-dominated), weakly oxidizing (\ce{N2}-dominated), or highly oxidizing (\ce{CO2}-dominated) atmospheres. The aerosol particle mean diameter is assumed to be 0.1 $\mu$m, and the \ce{H2S}/\ce{SO2} ratio of the surface emission is 0.5.  Other model parameters are tabulated in Table \ref{AtmosPara}.
Sulfur emission 2-orders-of-magnitude larger than current Earth's volcanic sulfur emission ($\sim3\times10^9$ S cm$^{-2}$ s$^{-1}$) leads to substantial aerosol opacities in the visible wavelengths in \ce{N2} and \ce{CO2} atmospheres, and sulfur emission comparable with current Earth's volcanic sulfur emission leads to substantial aerosol opacities in the visible wavelengths in \ce{H2} atmospheres. 
The wiggle in the concentration of sulfuric acid aerosols reflects the competition between two effects: more sulfur is available to be converted into sulfuric acid as the sulfur emission increases, but the atmosphere becomes more reducing and less oxidizing as the sulfur emission increases.
 }
 \label{Sulfur_S}
  \end{center}
\end{figure}

The main finding is that on terrestrial exoplanets having atmospheres ranging from reducing to oxidizing, the primary sulfur emission from the surface (e.g., \ce{H2S} and \ce{SO2}) is chemically short-lived. The sulfur emission leads to photochemical formation of elemental sulfur (\ce{S8}) and sulfuric acid (\ce{H2SO4}), which would condense to form aerosols if saturated in the atmosphere. In reducing atmospheres (e.g., \ce{H2} atmospheres), \ce{S8} aerosols are photochemically formed based on \ce{H2S} and \ce{SO2} emission; and in oxidized atmospheres (e.g., \ce{N2} and \ce{CO2} atmospheres), both \ce{S8} and \ce{H2SO4} aerosols may be formed (see Figure \ref{Sulfur_S}). In general, the higher the surface sulfur emission, the more aerosols exist in the atmosphere (see Figure \ref{Sulfur_S}).

As a result of photochemical production of elemental sulfur and sulfuric acid, terrestrial exoplanets with a habitable surface temperature (e.g., $270\sim320$ K) and substantial sulfur emission from the surface are likely to have hazy atmospheres. In this paper, we use ``hazy" to describe an atmosphere that has significant aerosol opacities at visible wavelengths (e.g., 500 nm). We find that even with an Earth-like surface sulfur emission, 1-bar \ce{H2} dominated atmospheres on habitable rocky exoplanets are hazy with \ce{S8} aerosols (see Figure \ref{Sulfur_S}). We also find that if the sulfur emission rate is $30\sim300$ times more than the Earth's current volcanic sulfur emission rate, photochemical \ce{S8} aerosols become optically thick at visible wavelengths in oxidized atmospheres including \ce{N2} and \ce{CO2} atmospheres (see Figure \ref{Sulfur_S}).

\begin{figure}[h]
\begin{center}
 \includegraphics[width=0.5\textwidth]{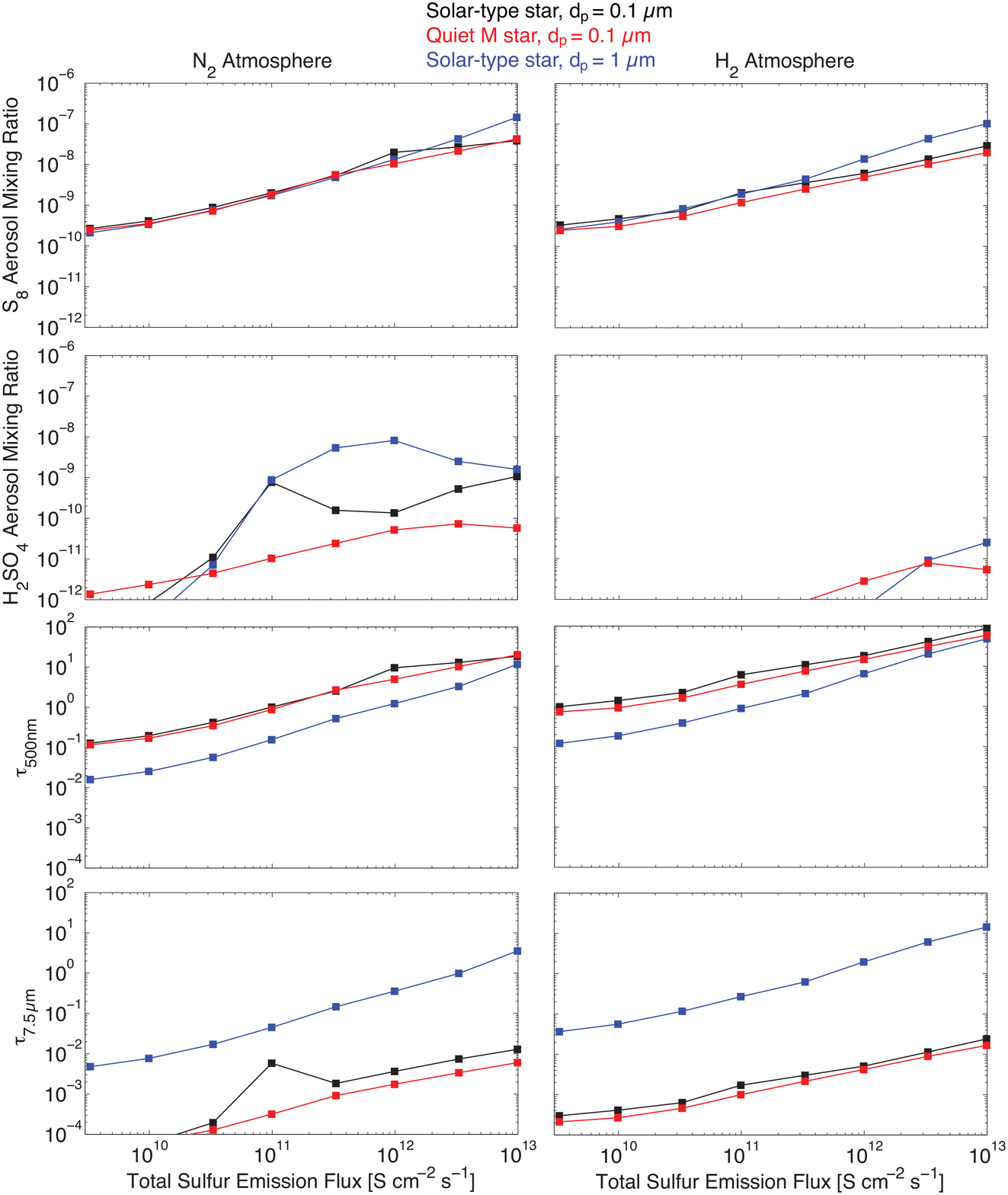}
 \caption{ 
 Aerosol mixing ratios and optical depths at the surface (1-bar pressure level) at 500 nm and 7.5 $\mu$m as a function of total sulfur emission rates, for an Earth-sized rocky planet orbiting a Sun-like star at 1 AU (black lines), a habitable planet around quiet M dwarf having effective temperature of 3100 K (red lines), and an Earth-sized rocky planet orbiting a Sun-like star at 1 AU with particle mean diameter of 1 $\mu$m (blue lines). The left column shows the case of weakly oxidizing \ce{N2} atmospheres, and the right column shows the case of reducing \ce{H2} atmospheres. The \ce{H2S}/\ce{SO2} ratio in the sulfur emission is 0.5 and other model parameters are tabulated in Table \ref{AtmosPara}. 
 We see that decreasing UV photon flux has little effect on the \ce{S8} formation, but results in a decrease of the amount of sulfuric acid aerosols, and therefore a decrease of MIR optical depth. We also see that particle diameter variation in $0.1\sim1$ $\mu$m has little effect on the chemical composition, but for similar mass abundance a larger particle size results in a smaller optical depth in the visible wavelengths and a larger optical depth in the MIR wavelengths.
 }
 \label{Sulfur_GMS}
  \end{center}
\end{figure}

\begin{figure}[h]
\begin{center}
 \includegraphics[width=0.5\textwidth]{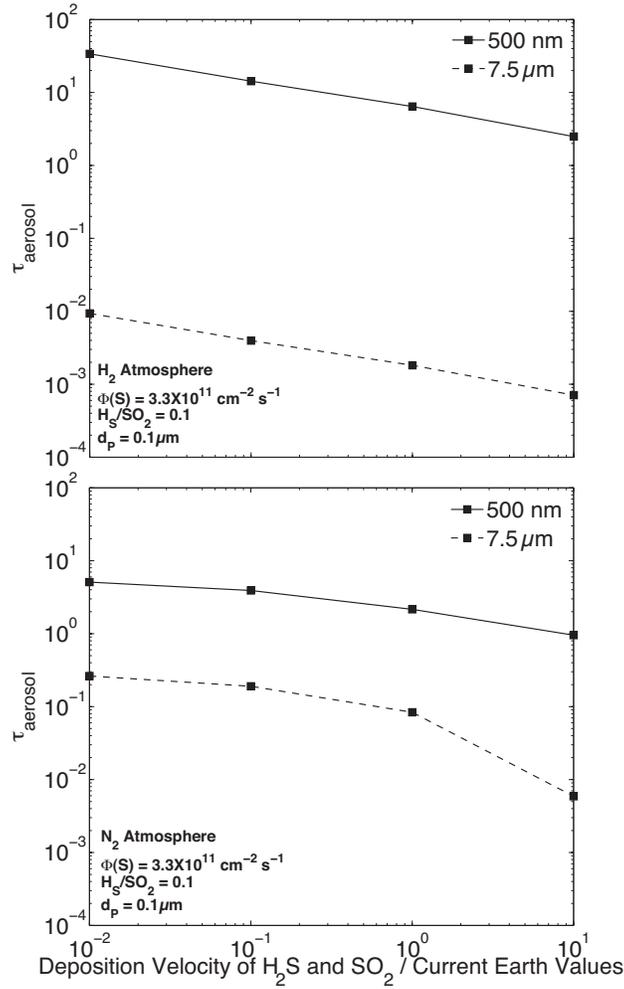}
 \caption{
 The relationship between the aerosol opacities (both \ce{S8} and \ce{H2SO4} aerosols) at 500 nm and 7.5 $\mu$m in the \ce{H2} and \ce{N2} atmospheres and the \ce{H2S} and \ce{SO2} dry deposition velocities. Model parameters are shown in the figure and tabulated in Table \ref{AtmosPara}.
 }
 \label{Sulfur_VB}
  \end{center}
\end{figure}

The key parameters that determine the aerosol opacity in the atmosphere are the surface sulfur emission rate, the dry deposition velocity, and the aerosol particle size. First, a higher surface sulfur emission rate leads to more sulfur and sulfate aerosols in anoxic atmospheres (e.g., Figure \ref{Sulfur_S} and Figure \ref{Sulfur_GMS}). Second, larger dry deposition velocities of \ce{H2S} and \ce{SO2} cause more rapid removal of these sulfur compounds from the atmosphere, which reduces the chance of converting them into condensable molecules (i.e., \ce{S8} and \ce{H2SO4}). Therefore, larger dry deposition velocities of \ce{H2S} and \ce{SO2} result in lower aerosol loading and aerosol opacities in the atmospheres, as shown in Figure \ref{Sulfur_VB}. Third, we find that the particle size has only secondary effects on the chemical composition of the atmosphere (i.e., by increasing the penetration of ultraviolet radiation), but has a primary effect on the aerosol optical depth. For mean particle diameter varying in the range of $0.1\sim1$ $\mu$m (i.e., typical particle sizes of photochemical aerosols on Earth (e.g., Seinfeld \& Pandis 2006) and Titan (e.g., Rages et al. 1983)), we do not see a notable variation in the yield of elemental sulfur, but we see an enhancement of \ce{H2SO4} production with large particles (Figure \ref{Sulfur_GMS}). Even with the same aerosol abundances, however, micron-sized particles cause lower opacities at the visible wavelengths and higher opacities in MIR compared with submicron-sized particles (Figure \ref{Sulfur_GMS}). 

\begin{figure}[h]
\begin{center}
 \includegraphics[width=0.5\textwidth]{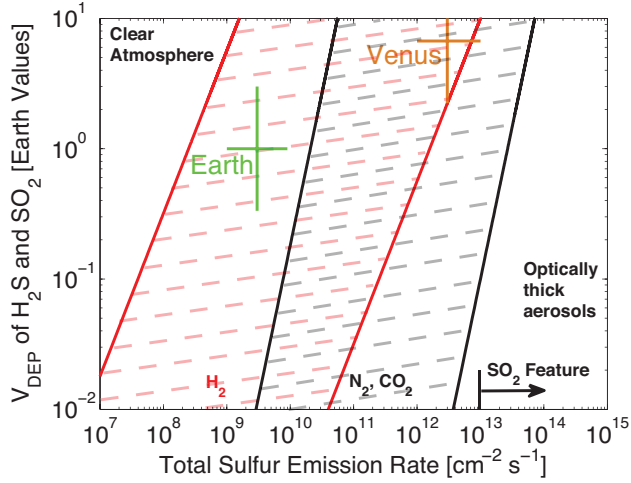}
 \caption{
Formation of optically thick aerosols in atmospheres on rocky exoplanets in the habitable zone of their host star as a result of surface sulfur emission and deposition.  The shaded areas are the parameter regime boundaries between a clear atmosphere and an optically thick atmosphere (defined as aerosol optical depth at 500-nm wavelength $\tau_{\rm 500 nm}>1$), for reducing (\ce{H2}) and oxidized (\ce{N2} and \ce{CO2}) atmospheres. 
The upper-left corner of the parameter regime (small sulfur emission rates, large deposition velocities) leads to clear atmospheres; whereas the lower-right corner of parameter regime (large sulfur emission rates, small deposition velocities) leads to optically thick aerosols in the atmosphere composed of sulfur (\ce{S8}) and sulfate (\ce{H2SO4}). 
The widths of the shaded boundary regime between clear atmospheres and optically thick atmospheres contain the uncertainties of: (1) the mean aerosol particle size ranging from 0.1 to 1 $\mu$m, (2) the \ce{H2S}/\ce{SO2} ratio of the sulfur emission ranging from 0.01 to 10 (i.e., more \ce{H2S} leads to thicker haze), (3) the spectral type of the host star ranging from G2V to M5, (4) the strength of vertical mixing in the atmosphere by eddy diffusion ranging from 0.1 to 10 times Earth's value, and (5) the surface temperature ranging from 270 to 320 K.
Earth and Venus are shown for a reference in the Solar System: Earth's volcanic sulfur emission and \ce{H2S} deposition velocity are plotted; and Venus' equivalent upward \ce{SO2} flux and \ce{SO2} deposition velocity at the altitude of 58 km are plotted (adapted from Krasnopolsky 2012). We note that the equivalent \ce{SO2} flux is a transfer rate across the 58-km altitude, and does not imply a surface emission rate.
The \ce{SO2} features at 7.5 $\mu$m and 20 $\mu$m requires a mixing ratio on the order of ppm to be spectrally significant, which corresponds to a sulfur emission flux of more than $10^{13}$ cm$^{-2}$ s$^{-1}$ due to rapid photochemical removal of \ce{SO2} in the atmosphere.
 }
 \label{Syn}
  \end{center}
\end{figure}

We capture the effects of the three key parameters on the aerosol opacity in anoxic atmospheres on terrestrial exoplanets by fitting the following power-law formula, i.e.,
\begin{equation}
\tau  = C\bigg(\frac{\Phi({\rm S})}{10^{11}\ {\rm cm}^{-2}\ {\rm s}^{-1}}\bigg)^a\bigg(\frac{V_{\rm DEP}}{V_{\rm DEP}({\rm Earth})}\bigg)^{-b}\bigg(\frac{d_{\rm P}}{0.1\ \mu{\rm m}}\bigg)^{-c}  , \label{EqnSyn}
\end{equation}
where $\tau$ is the vertical optical depth due to aerosols at 1 bar, $\Phi({\rm S})$ is the total sulfur emission rate, $V_{\rm DEP}/V_{\rm DEP} ({\rm Earth})$ is the dry deposition velocities of \ce{H2S} and \ce{SO2} with respect to current Earth values, $d_{\rm p}$ is the mean particle diameter of aerosols, $a$, $b$, and $c$ are positive numbers, and $C$ is a constant that covers other uncertainties. We have fit the empirical relation (\ref{EqnSyn}) through an extensive parameter exploration using photochemistry models (see Figure \ref{Sulfur_S} - \ref{Sulfur_VB} for examples) and determined the values of $C$, $a$, $b$ and $c$ for \ce{H2} dominated reducing atmospheres and for \ce{N2} and \ce{CO2} dominated oxidized atmospheres. We summarize graphically the parameter regime in which sulfur emission leads to a hazy atmosphere in Figure \ref{Syn}. Here we use $\tau _{500 \rm{nm}}$ and $\tau _{7.5 \rm{\mu m}}$ as the representatives for aerosol opacities at visible wavelengths and MIR wavelengths; due to the complex nature of the extinction cross sections of aerosol particles (Figure \ref{Cross_compare}), it is not practical to fold the full wavelength dependency into the empirical formula. Also, we find that it is always true that $\tau _{500 \rm{nm}}\sim0$ for mean particle diameter in the order of 10 $\mu$m and $\tau _{7.5 \rm{\mu m}}\sim0$ for mean particle diameter in the order of 0.1 $\mu$m.

For \ce{H2} atmospheres, and mean particle diameter $d_{\rm P}$ in the the range of $0.1\sim1$ $\mu$m,
\begin{equation}
\tau _{500 \rm{nm}} = 1\sim20\bigg(\frac{\Phi({\rm S})}{10^{11}\ {\rm cm}^{-2}\ {\rm s}^{-1}}\bigg)^{0.5}\bigg(\frac{V_{\rm DEP}}{V_{\rm DEP} ({\rm Earth})}\bigg)^{-0.4}\bigg(\frac{d_{\rm P}}{0.1\ \mu{\rm m}}\bigg)^{-0.6}  , \label{H2_Syn1}
\end{equation}
and for $d_{\rm P}$ in the the range of $1\sim10$ $\mu$m,
\begin{equation}
\tau _{7.5 \rm{\mu m}} = 0.1\sim1\bigg(\frac{\Phi({\rm S})}{10^{11}\ {\rm cm}^{-2}\ {\rm s}^{-1}}\bigg)^{0.7}\bigg(\frac{V_{\rm DEP}}{V_{\rm DEP} ({\rm Earth})}\bigg)^{-0.4}\bigg(\frac{d_{\rm P}}{1.0\ \mu{\rm m}}\bigg)^{-1.5}  . \label{H2_Syn2}
\end{equation}

For \ce{N2} and \ce{CO2} atmospheres, and $d_{\rm P}$ in the the range of 0.1 and 1 $\mu$m,
\begin{equation}
\tau _{500 \rm{nm}} = 0.1\sim3\bigg(\frac{\Phi({\rm S})}{10^{11}\ {\rm cm}^{-2}\ {\rm s}^{-1}}\bigg)^{0.7}\bigg(\frac{V_{\rm DEP}}{V_{\rm DEP} ({\rm Earth})}\bigg)^{-0.3}\bigg(\frac{d_{\rm P}}{0.1\ \mu{\rm m}}\bigg)^{-0.7}  , \label{N2_Syn1}
\end{equation}
and for $d_{\rm P}$ in the the range of $1\sim10$ $\mu$m,
\begin{equation}
\tau _{7.5 \rm{\mu m}} = 0.01\sim0.1\bigg(\frac{\Phi({\rm S})}{10^{11}\ {\rm cm}^{-2}\ {\rm s}^{-1}}\bigg)^{0.8}\bigg(\frac{V_{\rm DEP}}{V_{\rm DEP} ({\rm Earth})}\bigg)^{-0.5} \bigg(\frac{d_{\rm P}}{0.1\ \mu{\rm m}}\bigg)^{-1.6}. \label{N2_Syn2}
\end{equation}

The constant $C$ in equations (\ref{H2_Syn1} - \ref{N2_Syn2}) spans about one order of magnitude, which covers the variation of the following model inputs:
 \begin{itemize}
 \item The \ce{H2S}/\ce{SO2} ratio in the surface sulfur emission, ranging from 0.01 to 10;
 \item Temperature profiles deviating from the adopted temperature profile by $\pm30$ K that controls the mixing ratio of water vapor in the atmosphere by the cold trap;
 \item Stellar ultraviolet radiation received by the planet, ranging from the habitable zone of solar-like stars to the habitable zone of quiet M dwarfs with an effective temperature of 3100 K;
 \item Eddy diffusion coefficients ranging from 0.1 to 100 times the values of Earth's  atmosphere;
 \item Sulfur polymerization reaction rates (reactions \ref{SP0} - \ref{SP4}) ranging by one order of magnitude.
 \end{itemize}

To summarize, we find that the emission of \ce{H2S} and \ce{SO2} from the surface is readily converted into sulfur (\ce{S8}) and sulfate (\ce{H2SO4}) in anoxic atmospheres of terrestrial exoplanets. The photochemical sulfur and sulfate would condense to form aerosols if saturated in the atmosphere, which is likely to occur on a planet in the habitable zone of either a Sun-like star or a quiet M star.
The aerosol layer is optically thick at the visible and NIR wavelengths if the surface sulfur emission is comparable to Earth's volcanic sulfur emission in the \ce{H2} atmosphere, and more than $30\sim300$ times of the Earth's volcanic sulfur emission in other anoxic atmospheres, depending on the dry deposition velocities of sulfur compounds and particle size of the aerosols. 

\subsection{Spectral Features of \ce{SO2}, \ce{H2S}, and \ce{S8} and \ce{H2SO4} Aerosols}

\begin{figure}[h]
\begin{center}
 \includegraphics[width=0.5\textwidth]{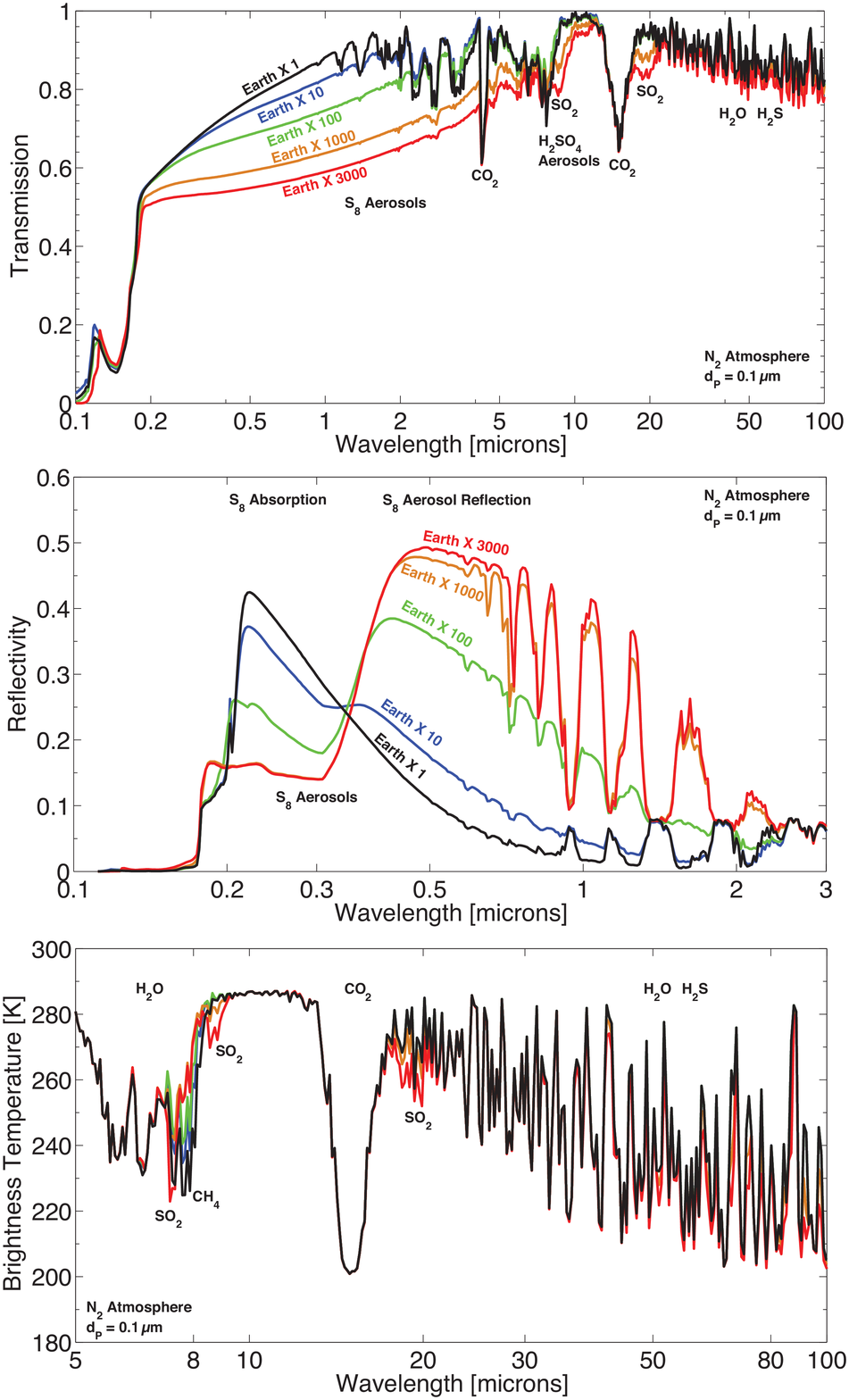}
 \caption{
 Transmission, reflection and thermal emission spectra of a terrestrial exoplanet with \ce{N2}-dominated atmosphere with various surface sulfur emission up to 3000 times Earth's current volcanic emission (labeled in colors).
The planet is an Earth-sized planet at the 1-AU orbit a Sun-like star, having surface temperature of 288 K.
 The \ce{H2S}/\ce{SO2} emission ratio is 0.5, the aerosol particle diameter is assumed to be 0.1 $\mu$m, and other model parameters are tabulated in Table \ref{AtmosPara}. The cross sections of \ce{S8} and \ce{H2SO4} aerosols are shown in Figure \ref{Cross_compare}.
 }
 \label{Spec}
  \end{center}
\end{figure}

The sulfur emission from surface shapes the spectra of terrestrial exoplanets at the visible and NIR wavelengths, mostly through the photochemical formation of \ce{S8} and \ce{H2SO4} aerosols. We use the model outputs from the photochemistry models to compute the transmission, reflection, and thermal emission spectra of a terrestrial exoplanet with various levels of sulfur emission, and show examples of the computed spectra in Figure \ref{Spec}. We see that submicron-sized \ce{S8} aerosols dominate the transmission and reflection spectra at wavelengths from visible up to 3 $\mu$m, if the sulfur emission is more than about two orders of magnitude higher than Earth's volcanic sulfur emission. In general, an atmosphere with high sulfur emission and therefore high aerosol loading generally exhibits a flat transmission spectrum (the \ce{H2O} features at NIR muted), and a high visible albedo (see Figure \ref{Spec}). Notably, \ce{S8} aerosols are purely reflective at 500 nm but absorptive at 300 nm. The absorption edge of \ce{S8} aerosols in 300 - 400 nm is evident in the reflection spectra for planets with enhanced sulfur emission (Figure \ref{Spec}), which is a potential diagnostic feature for \ce{S8} aerosols.

Although opaque at visible wavelengths, the atmospheres with enhanced sulfur emission are likely to be transparent in the MIR wavelengths ($\lambda>5$ $\mu$m). The spectral features of aerosols depend on their particle sizes, so we now consider two possibilities: if the particles are submicron-sized, the aerosol molecules have negligible cross sections at MIR (see Figure \ref{Cross_compare}); or if the particles are micron-sized, the falling velocity of aerosol particles is large enough to rapidly remove aerosols from the atmosphere, as implied by equation (\ref{H2_Syn2}) and equation (\ref{N2_Syn2}) that are applicable for micron-sized particles. Therefore in both cases the aerosol opacities at MIR are minimal even for very high sulfur emission rates (see Figure \ref{Spec} for examples of \ce{N2} atmospheres, and \ce{H2} atmospheres are qualitatively similar). The only exception, in which aerosols indeed affect MIR spectra, is the case of abundant \ce{H2SO4} aerosols. The main spectral effect of \ce{H2SO4} aerosols is absorption at MIR wavelengths ($5\sim10$ $\mu$m; Figure \ref{Spec}). However, the column-average mixing ratio of \ce{H2SO4} needs to be larger than 0.1 ppm in order to produce significant aerosol absorption at MIR. We find with numerical exploration that such a high abundance of \ce{H2SO4} aerosols is only possible in highly oxidizing \ce{CO2} dominated atmospheres without reducing gas emission (see Paper I for an example of such atmospheres). With reducing gas emission (i.e., \ce{H2} and \ce{CH4}), it is unlikely that \ce{H2SO4} mixing ratio exceeds 0.01 ppm in anoxic atmospheres for a wide range of sulfur emission rates (see Figure \ref{Sulfur_S}). In summary we expect the spectral effects of \ce{S8} and \ce{H2SO4} aerosols to be minimal at MIR for most cases.

We now turn to consider the direct spectral features of \ce{H2S} and \ce{SO2}. It has been previously proposed that \ce{H2S} and \ce{SO2} can be detectable on terrestrial exoplanets by their spectral features (Kaltenegger \& Sasselov 2010). However, our photochemistry models show that both \ce{H2S} and \ce{SO2} are chemically short-lived in the atmospheres, which implies that that substantial surface emission is required to maintain a detectable level of either \ce{H2S} or \ce{SO2} in the atmosphere. \ce{SO2} has diagnostic absorption features at 7.5 $\mu$m and 20 $\mu$m (see Figure \ref{Spec}). For these features to be detectable the mixing ratio of \ce{SO2} needs to be larger than 0.1 ppm, which corresponds to sulfur emission rates 1000 times more than current Earth's sulfur emission rates for \ce{H2}, \ce{N2}, and \ce{CO2} atmospheres (see Figure \ref{Sulfur_S}). The spectral feature of \ce{H2S} is the pseudo-continuum absorption at wavelengths longer than 30 $\mu$m, which coincides with the rotational bands of \ce{H2O}. We find that the only scenario in which \ce{H2S} may be directly detected is the case with extremely high sulfur emission rates (i.e., 3000 times higher than the current Earth's sulfur emission rate) on a highly desiccated planet without liquid water ocean so that there is no water vapor contamination. We therefore conclude that direct detection of \ce{H2S} and \ce{SO2} is tricky: they are chemically short-lived so that extremely large surface emission is required for a detectable mixing ratio in the atmosphere, and their spectral features may be contaminated by other gases in the atmosphere.

Finally, we suggest that the emission of sulfur compounds might be indirectly inferred by detecting sulfur and sulfate aerosols. Our numerical exploration reveals a monotonic relationship between the abundance of aerosols in the atmosphere and the emission rates of sulfur compounds (see Figure \ref{Sulfur_S}), and the composition of aerosols is correlated with the \ce{H2S}/\ce{SO2} ratio of the surface emission (see Figure \ref{N2_RATIO}).
A combination of featureless low atmospheric transmission (large planet radius viewed in transits) and high planetary albedo (large planetary flux at the visible wavelengths viewed in occultations) may establish the existence of aerosols in the atmosphere. In particular, elemental sulfur (\ce{S8}) aerosols are absorptive at wavelengths shorter than 400 nm and therefore might be identified by the absorption edge (see Figure \ref{Spec}). Sulfate aerosols (\ce{H2SO4}), if abundant in the atmosphere, lead to absorption features at the MIR wavelengths ($\lambda\sim5$ - 10 $\mu$m). However, none of these features are uniquely diagnostic of certain types of aerosols. The identification of aerosol composition, therefore, is by no means straightforward.
We learn from the Solar System exploration that the discriminating piece of information for aerosol identification comes from polarization of reflected stellar light. Historically, the bright clouds on Venus were identified to be mainly composed of \ce{H2SO4} droplets after the phase curve of the planet in polarized light had been observed (e.g., Young 1973; Hansen \& Hovenier 1974). We therefore postulate that aerosol identification on terrestrial exoplanets and the inference of surface sulfur emission might require observation of polarized reflected light as a function of planetary illumination phase.

\section{Discussion: Can \ce{H2S} be a Biosignature Gas?}

Biosignatures are gases in an exoplanet's atmosphere produced by life. In order to confirm a certain gas to be a plausible biosignature, one has to verify that the gas can accumulate in the atmosphere and that the amount of gas detected cannot be produced through plausible abiotic processes. It has been proposed and widely discussed that \ce{O2} (and its photolytic product \ce{O3}), \ce{N2O}, \ce{CH4} and \ce{CH3Cl} are exoplanet biosignatures (e.g., Sagan et al. 1993; Des Marais et al. 2002; Segura et al. 2005; Segura et al. 2007; Hu et al. 2012). Organosulfur compounds, such as dimethyl sulfide (DMS) and methanethiol (\ce{CH3SH}), have also been suggested to be biosignatures of the early Earth (Pilcher 2003) and anoxic exoplanets (Domagal-Goldman et al. 2011).

\ce{H2S} can be produced from several metabolic origins on Earth, and so is a candidate biosignature gas. Life on Earth can produce \ce{H2S} through sulfate reduction (when the environment is reduced) and sulfur disproportionation. Microorganisms can disproportionate sulfur compounds of intermediate oxidation states, including thiosulfate, sulfite, and elemental sulfur, into \ce{H2S} and sulfate (Finster 2008). For example, disproportionation of sulfite in the ocean is described by
\reaction{ 4 SO3^{2-} + H+ ->  3 SO4^{2-} + HS- , \label{SD}}
in which the Gibbs free energy released is 58.9 kJ mol$^{-1}$ sulfite. The sulfite reducers, including microorganisms in genus {\it Desulfovibrio} and {\it Desulfocapsa}, extract energy from the disproportionation (Kramer and Cypionka, 1989).

The effect of biotic \ce{H2S} production is the increase of the \ce{H2S}/\ce{SO2} ratio of the surface sulfur emission. If the \ce{H2S}/\ce{SO2} ratio in the volcanic sulfur emission is low (i.e., less than 0.1), sulfur disproportionation and sulfate reduction by life may increase the \ce{H2S}/\ce{SO2} ratio significantly, which may lead to a change in the redox input to the atmosphere and therefore the dominant aerosol species in the atmosphere, as suggested by Figure \ref{N2_RATIO}. Specifically, for a habitable terrestrial exoplanet having a weakly oxidizing \ce{N2} atmosphere, biotic production of \ce{H2S} in excess of the geological \ce{H2S} emission could result in a higher amount of \ce{S8} aerosols and a much lower amount of \ce{H2SO4} aerosols in the atmosphere compared with a planet without life. Although it is currently not possible to distinguish different types of aerosols, \ce{H2S} could be a biosignature gas in the long term.

The geological production of \ce{H2S}, and the consequent risk of a false positive mis-identification of geological \ce{H2S} for biological \ce{H2S}, will be a major obstacle of confirming \ce{H2S} to be a biosignature gas. Sulfur is believed to be present in the mantle of all terrestrial planets, and what determines the \ce{H2S}/\ce{SO2} ratio in the volcanic outgassing is the oxygen fugacity of the upper mantle, temperature of the location where magma degassing happens, water content in the conduit of magma, and gas content dissolved in the magma (e.g., Holland 1984; Kasting et al. 1985; Holland 2002; Burgisser \& Scaillet 2007). The current Earth volcanic emissions are oxidized, dominated by \ce{H2O}, \ce{CO2} and \ce{SO2} with minor contributions of \ce{H2}, \ce{CO} and \ce{H2S}. This volcanic gas composition is consistent with a magma buffered by the quartz-fayalite-magnetite (QFM) equilibrium, i.e., a relatively oxidized upper mantle (Holland 1984). As a global average, the volcanic \ce{H2S}/\ce{SO2} emission ratio on Earth is 0.1 (Holland 2002). However, if the mantle of a rocky exoplanet is much more reducing than that of the current Earth, significant geological source of \ce{H2S} can be expected (Holland 1984; Kasting 1993). As a result, it would be very hard to rule out a geological contribution to the \ce{H2S} emission flux by remote sensing.

In summary, although \ce{H2S} can be produced by energy-yielding metabolism, it is very unlikely to be a useful biosignature gas for three reasons. Firstly, \ce{H2S} itself is unlikely to be detectable directly by remote sensing because of its weak spectral features, and their contamination by the spectral features of water. Secondly, \ce{H2S} has a very short atmospheric lifetime, and so unrealistic emission rates are required to build up significant levels in any atmosphere. This second point could be overcome, in principle, by detecting \ce{S8} aerosols in an anoxic atmosphere, and discriminating them from \ce{H2SO4} aerosols. Discrimination between sulfur and sulfuric acid aerosols is not possible with current equipment, but may be possible in the future through analysis of reflected light. Thirdly, however, \ce{H2S} suffers from a significant false positive risk, as geological sources can also produce \ce{H2S}, and the ratio of \ce{H2S}/\ce{SO2} in geological emissions depends on mantle chemistry, the physical structure of the outgassing events, and the extent of surface reprocessing of vented sulfur gases. To infer that life was generating \ce{H2S} on a planet, this study shows that the observer would have to determine the \ce{S8}/\ce{H2SO4} aerosol ratio and have knowledge of the geological outgassing ratio of \ce{H2S}/\ce{SO2} and have knowledge of the surface chemistry that might modulate the primary outgassing rate. This seems an unreasonable requirement.

\section{Conclusion}

We studied the effect of \ce{H2S} and \ce{SO2} surface emission on anoxic atmospheres of terrestrial exoplanets. With a newly established one-dimensional photochemistry model that treats all relevant chemical reactions and photochemical processes of O, H, C, and S bearing species, as well as formation and sedimentation of sulfur and sulfate aerosols, we find that \ce{H2S} and \ce{SO2} gases emitted from surface are chemically short-lived in both reducing and oxidizing atmospheres. The sulfur emission results in photochemical production of elemental sulfur (\ce{S8}) and sulfuric acid (\ce{H2SO4}), which would condense to form aerosols if they are saturated in the atmosphere. 
For a planet in the habitable zone of a Sun-like star or a M star, Earth-like sulfur emission rates would result in optically thick aerosol layers in \ce{H2}-dominated atmospheres; and a sulfur emission rate 2-orders-of-magnitude higher than the Earth's volcanic sulfur emission rate would result in optically thick aerosol layers in \ce{N2} and \ce{CO2}-dominated atmospheres. 
The composition of the photochemically produced aerosols mostly depends on the redox state of the atmosphere: \ce{S8} aerosols are formed in the reducing atmospheres (e.g., \ce{H2} atmospheres), and both \ce{S8} and \ce{H2SO4} aerosols are formed in the oxidized atmospheres that could be both reducing and oxidizing (e.g., \ce{N2} and \ce{CO2} atmospheres). Based on extensive numerical simulations, we provide empirical formulae that show the dependency of the aerosol optical depth on the surface sulfur emission rates, the dry deposition velocities of sulfur compounds, and the aerosol particle sizes.

We find that direct detection of \ce{H2S} and \ce{SO2} is unlikely due to the rapid photochemical conversion from \ce{H2S} and \ce{SO2} to elemental sulfur and sulfuric acid in atmospheres having a wide range of redox powers. For a terrestrial exoplanet with sulfur emitted from the surface at an enhanced rate, it is likely that at visible wavelengths the planet's atmosphere appears to be opaque due to the aerosol loading and that the planet has high visible albedo. However, for Earth-like planets with 1-bar atmospheres ranging from reducing to oxidizing, we find the effect of photochemical sulfur and/or sulfate aerosols in the MIR wavelengths is minimal, because micron-sized particles that interact with MIR photons have large gravitational settling velocities and therefore short atmospheric lifetime. Finally, as the aerosol composition is tightly related to the ratio of the \ce{H2S} versus \ce{SO2} emission, although direct detection of \ce{H2S} and \ce{SO2} by their spectral features is unlikely, their existence might be inferred by observing aerosol-related features in reflected light with future generation space telescopes.

\acknowledgments

We thank Shuhei Ono and Jenny Suckale for helpful discussions about the sulfur-bearing gases release by the volcanism. We thank Feng Tian for providing the refractive index data of \ce{S8} aerosols, and Antigona Segura for discussions regarding \ce{CO2} atmospheres. We thank James Kasting for helpful suggestions about the photochemical model. We thank Linda Elkins-Tanton for helpful suggestions on the mantle degassing. We thank Vikki Meadows for discussions about detecting clouds on Venus. We thank the anonymous referee to the improvement of the manuscript. RH is supported in part by the NASA Earth and Space Science Fellowship (NESSF/NNX11AP47H).

\clearpage

\clearpage

\clearpage

\clearpage

\clearpage

\clearpage

\clearpage

\clearpage

\clearpage

\begin{thebibliography}{}

\bibitem[]{} Allard, F., Hauschildt, P. H., Alexander, D. R., Starrfield, S., 1997, Annu. Rev. Astron. Astrophys., 35, 137







\bibitem[]{} Batalha, N. M., Borucki, W. J., Bryson, S. T., et al., 2011, ApJ, 729, 27

\bibitem[]{} Bean, J., Miller-Ricci, E., Homeier, D., 2010, Nature, 468, 669

\bibitem[]{} Berta, Z. K., Charbonneau, D., D\'esert, J.-M., et al., 2012, ApJ, 747, 35




\bibitem[]{} Borysow, A., 2002, A\&A, 390, 779

\bibitem[]{} Burgisser, A., Scaillet, B., 2007, Nature, 445, 194

\bibitem[]{} Carlson, R. W., Kamp, L. W., Baines, K. H., Pollack, J. B., Grinspoon, D. H., Encrenaz, Th., Drossart, P., Tayler, F. W., 1993, Planet. Space Sci., 41, 477

\bibitem[]{} Croll, B., Albert, L., Jayawardhana, R., et al., 2011, ApJ, 736, 78








\bibitem[]{} Demory, B.-O., Gillon, M., Seager, S., 2012, ApJ, 751, 28

\bibitem[]{} de Mooij, E. J. W., Brogi, M., de Kok, R. J., et al., 2012, A\&A, 538, A46

\bibitem[]{} D\'esert, J.-M., Bean, J., Miller-Ricci Kempton, E., et al., 2011, ApJ, 731, L40

\bibitem[]{} Des Marais, D. J.,  et al., 2002, Astrobiology, 2, 153

\bibitem[]{} Domagal-Goldman, S. D., Meadows, V. S., Claire, M. W., Kasting, J. F., 2011, Astrobiology, 11, 1

\bibitem[]{} Ehrenreich, D., Bourrier, V., Bonfils, X., 2012, A\&A, 547, 18




\bibitem[]{} Farquhar, J., Bao, H., Thiemans, M., 2000, Science, 289, 756

\bibitem[]{} Finster, K., 2008, Journal of Sulfur Chemistry, 29, 281





\bibitem[]{} Halevy, I., Zuber, M. T., Schrag, D. P., 2007, Science, 318, 1903


\bibitem[]{} Hansen, J. E., Hovenier, J. W., 1974, Journal of Atmospheric Sciences, 31, 1137

\bibitem[]{} Hauglustaine, D. A., Granier, C., Brasseur, G. P., M\'egie, G., 1994, Journal of Geophysical Research, 99, 1173



\bibitem[]{} Holland, H. D., 1984, Chemical Evolution of the Atmosphere and Oceans, Princeton University Press, Princeton, New Jersey

\bibitem[]{} Holland, H. D., 2002, Geochim. Cosmichim. Acta, 66, 3811



\bibitem[]{} Holton, J. R., 1986, Journal of Geophysical Research, 91, 2681

\bibitem[]{} Hu, R., Seager, S., Bains, W., 2012, ApJ, 761, 166

\bibitem[]{} Jones, A. D., 1976, J. Quant. Spectrosc. Radiat. Transfer, 16, 1017


\bibitem[]{} Kaltenegger, L., Sasselov, D., 2010, ApJ, 708, 1162

\bibitem[]{} Kasting, J. F., 1990, Origins of Life and Evolution of the Biosphere, 20, 199

\bibitem[]{} Kasting, J. F., Holland, H. D., Pinto, J. P., 1985, Journal of Geophysical Research, 90, 10497

\bibitem[]{} Kasting, J. F., Zahnle, K. J., Pinto, J. P., Young, A. T., 1989, Origins of Life and Evolution of the Biosphere, 19, 95

\bibitem[]{} Kasting, J. F., Whitmire, D. P., Reynolds, R. T., 1993, Icarus, 101, 108

\bibitem[]{} Kasting, J. F., Catling, D., 2003, ARA\&A, 41, 429



\bibitem[]{} Kramer, M., Cypionka, H., 1989, Arch. Microbiol., 151, 232



\bibitem[]{} Krasnopolsky, V. A., 2012, Icarus, 218, 230






\bibitem[]{} Lelieveld, J., Roelofs, G. J., Ganzeveld, L., Feichter, J., Rodhe, H., 1997, Phil. Trans. Roy. Soc. Lond., B352, 149

\bibitem[]{} Lodders, K., 2003, ApJ, 591, 1220



\bibitem[]{} Lyons, J. R., 2008, Journal of Sulfur Chemistry, 29, 269

\bibitem[]{} Madhusudhan, N., Seager, S., 2009, ApJ, 707, 24                  


\bibitem[]{} Massie, S. T., Hunten, D. M., 1981, Journal of Geophysical Research, 86, 9859


\bibitem[]{} Meyer, B., 1976, Chemical Reviews, 76, 368

\bibitem[]{} Miller-Ricci, E., Seager, S., Sasselov, D., 2009, ApJ, 690, 1056



\bibitem[]{} Moses, J. I., Zolotov, M. Y., Fegley, B., 2002, Icarus, 156, 76









\bibitem[]{} Palmer, K. F., Williams, D., 1975, Applied Optics, 14, 208

\bibitem[]{} Pavlov, A. A., Kasting, J. F., 2002, Astrobiology, 2, 27


\bibitem[]{} Pierrehumbert, R., Gaidos, E., 2011, ApJ, 734, L13

\bibitem[]{} Pilcher, C. B., 2003, Astrobiology, 3, 471


\bibitem[]{} Rages, K., Pollack, J. B., Smith, P. H., 1983, JGR, 88, 8721

\bibitem[]{} Rothman, L. S., et al., 2009, Journal of Quantitative Spectroscopy \& Radiative Transfer, 110, 533

\bibitem[]{} Sander, S. P., et al. 2011, Chemical Kinetics and Photochemical Data for Use in Atmospheric Studies, Evaluation Number 17, JPL Publication 10-6

\bibitem[]{} Sagan, C., Thompson, W. R., Carlson, R., Gurnett, D., Hord, C., 1993, Nature, 365, 715

\bibitem[]{} Sasson, R., Wright, R., Arakawa, E. T., Khare, B. N., Sagan, C., 1985, Icarus, 64, 368

\bibitem[]{} Seager, S., Sasselov, D. D., 2000, ApJ, 537, 916


\bibitem[]{} Seager, S., Whitney, B. A., Sasselov, D. D., 2000, ApJ, 540, 504

\bibitem[]{} Sehmel, G. A., 1980, Atmospheric Environment, 14, 983

\bibitem[]{} Seinfeld, J. H., Pandis, S. N., 2006, Atmospheric Chemistry and Physics - From Air Pollution to Climate Change, 2nd Edition, Published by Wiley, New Jersey


\bibitem[]{} Segura, A., Kasting, J. F., Meadows, V., Cohen, M., Scalo, J., Crisp, D., Butler, R. A. H., Tinetti, G., 2005, Astrobiology, 5, 706

\bibitem[]{} Segura, A., Meadows, V. S., Kasting, J. F., Crisp, D., Cohen, M., 2007, A\&A, 472, 665

\bibitem[]{} Smith, M. D., 1998, Icarus, 132, 176






\bibitem[]{} Tian, F., Claire, M. W., Haqq-Misra, J. D., Smith, M., Crisp, D. C., Catling, D., Zahnle, K., Kasting, J. F., 2010, Earth and Planetary Science Letters, 295, 412



\bibitem[]{} Toon, O. B., McKey, C. P., Ackerman, T. P., 1989, JGR, 94, 16287


\bibitem[]{} Van de Hulst, H. C., 1981, Light Scattering by Small Particles, New York: Dover


\bibitem[]{} Wallace, P. J., Edmonds, M., 2011, Reviews in Mineralogy \& Geochemistry, 73, 215

\bibitem[]{} Watts, S. F., 2000, Atmospheric Environment, 34, 761

\bibitem[]{} Wordsworth, R., 2012, Icarus, 219, 267

\bibitem[]{} Young, A. T., 1973, Icarus, 18, 564

\bibitem[]{} Yung, Y. L., DeMore, W. B., 1982, Icarus, 51, 199

\bibitem[]{} Yung, Y. L., DeMore, W. B., 1999, Photochemistry of Planetary Atmospheres, Oxford University Press


\bibitem[]{} Zahnle, K., Claire, M., Catling, D., 2006, Geobiology, 4, 271


\bibitem[]{} Zahnle, K., Marley, M. S., Freedman, R. S., Lodders, K., Fortney, J., 2009, ApJ, 701, L20

\bibitem[]{} Zhang, X., Liang, M. C., Mills, F. P., Belyaev, D. A., Yung, Y. L., 2012, Icarus, 217, 714

\end{thebibliography}
\end{document}